\documentclass[aps,prd,reprint,superscriptaddress,longbibliography,showpacs,nofootinbib]{revtex4-2}

\usepackage{graphicx}
\usepackage{svg}
\usepackage{dcolumn}
\usepackage{bm}

\usepackage[figurename=Figure]{caption}
\usepackage{ragged2e}
\DeclareCaptionJustification{plain}{\justifying}
\usepackage[percent]{overpic}
\usepackage{float}    
\usepackage{verbatim} 
\usepackage{amsmath}  
\usepackage{bbold}
\usepackage{upgreek} 
\usepackage{amssymb}  
\usepackage{svg}
\usepackage{enumitem} 
\usepackage{latexsym,epsfig}
\usepackage{subcaption}
\usepackage{stackrel}
\usepackage{mathtools}

\usepackage[colorlinks=true,breaklinks=true,allcolors=blue]{hyperref}
\usepackage[capitalize]{cleveref}
\usepackage{lipsum}
\usepackage{dsfont}
\usepackage{coffeestains}
\usepackage{tabularx}
\usepackage{placeins}

\usepackage[scr=boondox]{mathalfa}
 
\usepackage[scr=boondox]{mathalfa}

\newcommand{\be}{\begin{equation}}
	\newcommand{\ee}{\end{equation}}
\newcommand{\bea}{\begin{equation}\begin{aligned}}
		\newcommand{\eea}{\end{aligned}\end{equation}}
\newcommand{\ben}{\begin{enumerate}}
	\newcommand{\een}{\end{enumerate}}

\DeclareDocumentCommand{\nint}{ O{} O{} m }{\ensuremath{ \int_{\mbox{\scriptsize $#1$}}^{\mbox{\scriptsize$#2$}}\!\!\! \mbox{\small $\,\mathrm{d}#3$\! }}}

\usepackage{tikz,bm} 
\usepackage{circuitikz}
\usetikzlibrary{decorations.markings}
\usetikzlibrary{decorations.pathreplacing,calligraphy}

\definecolor{mycolor}{rgb}{1,0.2,0.3}
\definecolor{brightgreen}{rgb}{0.4, 1.0, 0.0}
\definecolor{britishracinggreen}{rgb}{0.0, 0.26, 0.15}
\definecolor{cadmiumgreen}{rgb}{0.0, 0.42, 0.24}
\definecolor{ceruleanblue}{rgb}{0.16, 0.32, 0.75}
\definecolor{darkelectricblue}{rgb}{0.33, 0.41, 0.47}
\definecolor{darkpowderblue}{rgb}{0.0, 0.2, 0.6}
\definecolor{darktangerine}{rgb}{1.0, 0.66, 0.07}
\definecolor{emerald}{rgb}{0.31, 0.78, 0.47}
\definecolor{palatinatepurple}{rgb}{0.41, 0.16, 0.38}
\definecolor{pastelviolet}{rgb}{0.8, 0.6, 0.79}
\begin{document}
	
	\preprint{APS/123-QED}
	
	\title{Long-Range Pairing in the Kitaev Model: Krylov Subspace Signatures}
	\author{Rishabh Jha}
	\email{rishabh.jha@uni-goettingen.de}
	\affiliation{%
		Institute for Theoretical Physics, Georg-August-Universit\"{a}t G\"{o}ttingen, Friedrich-Hund-Platz 1, 37077 G\"{o}ttingen, Germany
	}
	\author{Heiko Georg Menzler}
	\email{heiko.menzler@uni-goettingen.de}
	\affiliation{%
		Institute for Theoretical Physics, Georg-August-Universit\"{a}t G\"{o}ttingen, Friedrich-Hund-Platz 1, 37077 G\"{o}ttingen, Germany
	}%
	%


	\begin{abstract}

Krylov subspace methods quantify operator growth in quantum many-body systems through Lanczos coefficients that encode how operators spread under time evolution. Although these diagnostics were originally motivated by questions of chaos and integrability, quadratic fermionic Hamiltonians are often expected to exhibit trivial Lanczos structure. Here we show that, in the long-range Kitaev chain, Lanczos coefficients generated from local boundary operators sharply diagnose whether the lowest excitation gap is controlled by boundary-localized or bulk-extended modes. We introduce the \textit{Krylov staggering parameter} for the Lanczos coefficients. In the short-range Kitaev chain with balanced hopping and pairing, we derive analytically for arbitrary system size (valid in the thermodynamic limit) and show that this quantity is exactly constant and its sign cleanly distinguishes the topological phase with Majorana edge modes from the trivial phase. Away from that limit, long-range couplings and pairing-hopping imbalance deform the simple flat structure and analytical control is lost, nevertheless, we show that the sign pattern of the diagnostic still tracks whether the lowest excitation gap is controlled by boundary modes or by bulk excitations. These results are enabled by an exact single-particle operator Lanczos algorithm, as derived in this work, which reduces the recursion from exponentially large operator space to a finite-dimensional linear problem and achieves machine precision for chains of hundreds of sites. Krylov diagnostics thus emerge as practical probes of boundary-versus-bulk low-energy physics in topological superconductors with broken $U(1)$ symmetry and algebraically decaying couplings.

	\end{abstract}
	
	\maketitle
	

	
	\section{Introduction}
	\label{sec:introduction}

	Operator growth under Heisenberg time evolution provides a concrete route to understanding equilibration and the spread of quantum information in many-body systems. While out-of-time-ordered correlators (OTOCs) quantify scrambling directly~\cite{Sekino2008,Shenker2014,Hosur2016,Swingle2018}, an alternative and complementary viewpoint is obtained by expanding the evolving operator in a Krylov basis generated by repeated commutators with the Hamiltonian. Implemented via the Lanczos algorithm~\cite{Lanczos1950,Viswanath1994, capizzi2026universalpropertiesmanybodylanczos}, this procedure tridiagonalizes the Liouvillian and produces a sequence of Lanczos coefficients $\{b_n\}$ that encodes how the operator explores Krylov space. For Hermitian seeds (physical observables), the diagonal coefficients vanish identically, so the dynamics reduces to a purely off-diagonal, tri-diagonal, tight-binding-like problem on the Krylov chain.
	
	The recent surge of interest in Krylov/Lanczos diagnostics was driven in part by the ``universal operator growth hypothesis'' (UOGH)~\cite{Parker2019, rabinovici2025krylovcomplexity, Nandy2025Jun, Baiguera2026Feb}, which proposed that for chaotic Hamiltonians, any simple local operator seed exhibits a robust, near-universal regime of linear-in-$n$ growth of $b_n$. However, it is now clear that $b_n$-profiles cannot be used as a standalone discriminator of quantum chaos: integrable systems may be consistent with UOGH-like growth~\cite{Bhattacharjee2022}, and rapid scrambling can produce similar Lanczos profiles even though scrambling itself is necessary but not sufficient for chaos~\cite{Xu2020Apr,Dowling2023Nov}. Even more sharply, strictly quadratic models (and hence exactly solvable and non-ergodic) can be tuned to display a wide range of ``universal-looking'' Lanczos behavior, underscoring that $\{b_n\}$ alone do not classify integrability, ergodicity, or chaos~\cite{kehrein_krylov}.

	A particularly widespread expectation, as articulated already in the original UOGH discussion, is that for quadratic fermionic Hamiltonians the Lanczos structure is essentially trivial (often summarized as ``$b_n$ becomes constant'' for simple free models)~\cite{Parker2019}. This expectation is plausible: free systems do not thermalize in the generic Eigenstate Thermalization Hypothesis (ETH) sense and do not exhibit many-body chaos. Yet it is also potentially misleading. Quadratic Hamiltonians can host nontrivial boundary physics, localization properties, and gap mechanisms that are not captured by any single ``universal'' growth law for $\{b_n\}$.
	These observations shift the central question: can Lanczos coefficients encode \emph{other} physically sharp distinctions beyond integrability versus chaos? Specifically, can Krylov diagnostics resolve whether the lowest excitation gap is controlled by boundary-localized or bulk-extended modes?
	Answering this quantitatively poses a technical challenge: standard many-body Krylov implementations suffer numerical instabilities at large recursion depths, limiting accessible system sizes and making it difficult to distinguish genuine physics from finite-precision artifacts. The answer to both questions is affirmative, as we demonstrate through an exact single-particle formulation that achieves machine precision for chains with hundreds of sites.

	In parallel, long-range interacting quantum systems have revealed qualitatively new behavior compared to short-range models~\cite{Defenu2023}. Power-law couplings $\propto r^{-\alpha}$ (where $r$ is the spatial separation) can lead, for example, to area-law violations~\cite{Koffel2012,Vodola2014Oct,Eisert2010}, hybrid exponential-algebraic correlations~\cite{Vodola2015Dec,FossFeig2015,Gong2014}, modified criticality and conformal-symmetry breaking~\cite{Vodola2015Dec,Maghrebi2015,Gong2015}, and gapped phases without conventional bulk-gap-closure transitions~\cite{Vodola2015Dec,Gong2015}. A paradigmatic fermionic setting is the long-range Kitaev chain~\cite{Vodola2014Oct,Vodola2015Dec}, where both hopping and pairing decay algebraically. The short-range Kitaev chain famously supports boundary Majorana zero modes and realizes a topological superconductor in class BDI~\cite{Kitaev2001,Altland1997Jan,Read2000,Alicea2012,Beenakker2013,Sarma2015}; in that limit it maps to the transverse-field Ising chain via Jordan-Wigner~\cite{Lieb1961,Kitaev2001}. For long-range couplings, the Jordan-Wigner map becomes nonlocal, but the fermionic model remains quadratic and exactly solvable. Prior work established that the long-range Kitaev chain exhibits a rich phase structure, including regimes with massless edge modes at sufficiently large $\alpha$ and massive edge modes for $\alpha<1$ (depending on hopping-pairing imbalance), together with hybrid correlation profiles even in gapped phases~\cite{Vodola2015Dec}.
	
	Returning to the central question posed above, this paper asks whether Krylov subspace diagnostics can capture the interplay of (i) long-range pairing, (ii) $U(1)$ breaking to fermion parity, and (iii) the competition between boundary-localized and bulk-extended low-energy excitations \emph{within a quadratic model}. Concretely: do edge-dominated and bulk-dominated gap regimes leave distinct and robust signatures in Lanczos data generated from local boundary operators? 
	Addressing this in the long-range Kitaev chain is nontrivial precisely because the model remains free for all $\alpha$, so any observed structure must arise from the structure of the eigenmodes and their spatial localization rather than from chaos or thermalization.

Our main results are twofold. First, for the \emph{short-range} Kitaev chain at the hopping-pairing-balanced point, we derive an exact closed-form expression for the entire Lanczos recursion generated by the boundary Majorana seed $\gamma_1$: for every recursion depth $n$ and every system size $N$ (all the way to the thermodynamic limit), the off-diagonal coefficients $b_n$ are constant and can be obtained exactly, while all diagonal coefficients vanish, $a_n=0$ (Appendix~\ref{app:srk_exact}). Equivalently, the \emph{Krylov staggering parameter} $\eta_n \equiv \ln(b_{2n-1}/b_{2n})$ is exactly $n$-independent, with no finite-size corrections and no approximations. This result is not merely a compact formula for Lanczos coefficients: the balanced short-range Kitaev chain in Majorana form is exactly the SSH chain, and the Krylov recursion generated by the boundary Majorana seed preserves this alternating-bond matrix structure. We find, the sign of \(\eta_n\) is an exact analytic topological order parameter. In particular, \(\eta_n<0\) if and only in the topological phase of the physical Kitaev chain with Majorana zero modes, while \(\eta_n>0\) if and only in the trivial phase, thereby recovering Kitaev's phase boundary from a purely Krylov-recursion argument without invoking band topology, winding numbers, or bulk-gap closure.
		
Second, for the \emph{long-range} Kitaev chain, the exact flat short-range structure is deformed by power-law couplings, so $\eta_n$ is no longer constant in $n$. The crucial lesson from the short-range theorem, however, survives: each value of $\eta_n$ continues to encode the local SSH-type bonding character of the Krylov chain at recursion depth $n$, with $\eta_n<0$ corresponding to a locally topological strong-bond regime and $\eta_n>0$ to a locally trivial weak-bond regime. This motivates a new diagnostic principle for the deformed problem: not the constancy of $\eta_n$, but its \emph{sign structure} across $n$. We show empirically across the full long-range/short-range and pairing imbalance phase diagram that sign changes of $\eta_n$ correlate robustly with regimes in which the lowest excitation gap is controlled by boundary-localized modes, whereas a fixed positive sign correlates with bulk-dominated trivial regimes. The key technical ingredient enabling both statements is an exact single-particle Lanczos construction: because the commutator algebra closes on operators linear in Majorana modes, Heisenberg evolution is governed by the finite-dimensional Hermitian generator $L_{\mathrm{sp}}=i\mathcal{H}_M$, reducing the Krylov problem from exponentially large operator space to a $2N$-dimensional linear problem and allowing machine-precision calculations for chains with hundreds of sites. In this way, Krylov subspace dynamics become a practical and quantitatively sharp probe of whether the low-energy gap is edge-dominated or bulk-dominated even in a quadratic, non-chaotic system.

    The underlying mechanism is rooted in a bipartite Krylov structure: for a real boundary seed and real antisymmetric Majorana Hamiltonian $\mathcal{H}_M$, successive Krylov vectors alternate between real and purely imaginary subspaces, so $L_{\mathrm{sp}}^2$ acts separately on odd and even sectors.
	When a low-energy edge mode dominates the seed overlap, the Krylov recursion produces an imbalance between the odd and even subsequences as recursion depth increases, which leads to robust sign changes in $\eta_n$.
	When the lowest scale is set by bulk modes, the two subsequences remain more evenly matched and $\eta_n$ retains a fixed sign.
	While Ref.~\cite{Caputa2023Jan} studied topological transitions using state (spread) Krylov complexity, and Ref.~\cite{Chakrabarti_2025} analyzed monitored Su–Schrieffer–Heeger (SSH) dynamics in Krylov space within a similar state-spread framework (including connections to quantum Fisher information), we instead use the operator Lanczos algorithm and derive an exact single-particle formulation to extract Lanczos coefficients for boundary seeds, which directly resolve the edge–bulk gap competition.
	In this way, operator dynamics captured in Krylov subspace provide a practical probe of boundary-versus-bulk gap control, despite the model being quadratic.

	The remainder of this paper is organized as follows. In Sec.~\ref{sec:model} we introduce the long-range Kitaev Hamiltonian and its BdG and Majorana formulations, establish closure of the commutator algebra on Majorana-linear operators, and derive the exact Krylov structure of the hopping-pairing-balanced short-range case in the thermodynamic limit (Sec.~\ref{subsec:srk:anchor}), which serves as an analytic anchor for the long-range analysis. Sec.~\ref{sec:algorithm} derives the single-particle operator Lanczos algorithm, derives the associated tridiagonal representation of the single-particle Liouvillian, and proves that the diagonal Lanczos coefficients vanish identically for Hermitian seed operators (Appendix~\ref{app:alg_an_zero}). In Sec.~\ref{sec:physics of Krylov Structure for Long-Ranged Pairing in Majorana Basis}, we define edge- versus bulk-gap regimes and introduce the Krylov staggering parameter diagnostic. Sec.~\ref{sec:results} presents Lanczos-coefficient data and joint phase diagrams showing quantitative agreement between the gap-based and Krylov-based classifications. Sec.~\ref{sec:conclusion} discusses implications, including disordered long-range settings~\cite{Cinnirella2025}, interacting generalizations, and experimental prospects in cold-atom and trapped-ion platforms~\cite{Richerme2014,Jurcevic2014}.
	
	\section{Model}
	\label{sec:model}
	
	\subsection{Hamiltonian and parameters}
	\label{subsec:model_hamiltonian}
	
	We consider a chain of $N$ sites with open boundary conditions throughout.
	The degrees of freedom are spinless fermions with annihilation (creation) operators $c_j$ ($c_j^\dagger$) obeying the canonical anticommutation relations
	$\{c_j,c_k^\dagger\}=\delta_{jk}$ and $\{c_j,c_k\}=0$.
	The long-range Kitaev Hamiltonian is
	\begin{equation}
		H_{\mathrm{LRK}}
		=\sin \theta \sum_{1\le i < j\le N} \frac{c_i^{\dagger} c_j+(1+\epsilon) c_i c_j+\mathrm{h.c.}}{|i-j|^\alpha} +2 \cos \theta \sum_{i=1}^N n_i,
		\label{eq:hamiltonian_long_range_kitaev}
	\end{equation}
	where $n_i=c_i^\dagger c_i$.
	The exponent $\alpha>0$ controls the algebraic decay of both hopping and pairing amplitudes.
	The angle $\theta$ interpolates between the long-range kinetic and pairing sector (prefactor $\sin\theta$) and the on-site term (prefactor $2\cos\theta$), which plays the role of a chemical potential.
	Clearly, due to the presence of pairing terms, this model does not respect particle number conservation.
	The parameter $\epsilon$ quantifies an imbalance between hopping and pairing strengths: $\epsilon=0$ corresponds to equal amplitudes, while $\epsilon\neq 0$ biases the pairing relative to hopping and can qualitatively affect edge mode properties. The special case of $\epsilon = -1$ corresponds to the case of a simple tight-binding model with particle number conservation.
	
	In the short-range limit $\alpha\to\infty$, only nearest neighbor couplings survive and Eq.~\eqref{eq:hamiltonian_long_range_kitaev} reduces to the standard Kitaev chain with hopping $t=\sin\theta$, pairing $\Delta=(1+\epsilon)\sin\theta$, and chemical potential $\mu=-2\cos\theta$ (up to an overall sign convention for $H_{\mathrm{LRK}}$).
	
	\subsection{Spin-duality and Jordan-Wigner strings}
	\label{subsec:model_duality}
	
	In the short-range limit, the Kitaev chain maps to the transverse field Ising chain through the Jordan-Wigner transformation \cite{Chhajed2021Nov}.
	Introducing Pauli operators $\sigma_j^{x,y,z}$ and $\sigma_j^\pm=(\sigma_j^x\pm i\sigma_j^y)/2$, we define
	\begin{equation}
		c_j=\left(\prod_{\ell<j}\sigma_\ell^z\right)\sigma_j^-,
		\quad
		c_j^\dagger=\left(\prod_{\ell<j}\sigma_\ell^z\right)\sigma_j^+,
		\quad
		\sigma_j^z=2n_j-1.
		\label{eq:JW}
	\end{equation}
	For nearest neighbor couplings, the Jordan-Wigner strings $\prod_{\ell<j}\sigma_\ell^z$ cancel between adjacent sites, yielding a local spin Hamiltonian (the Ising chain in a transverse field).
	
	For finite $\alpha$, Eq.~\eqref{eq:hamiltonian_long_range_kitaev} contains fermionic bilinears between distant sites.
	Under Eq.~\eqref{eq:JW}, terms such as $c_i^\dagger c_j$ and $c_i c_j$ acquire nonlocal string operators extending over the interval $(i,j)$.
	Therefore, the long-range fermionic Hamiltonian \eqref{eq:hamiltonian_long_range_kitaev} is not simply equivalent to a spin Hamiltonian with pairwise long-range $\sigma_i^x\sigma_j^x$ couplings, even though the short-range limit recovers the familiar Ising-Kitaev correspondence \cite{Vodola2015Dec}.
	This distinction is crucial for what follows: the fermionic model \eqref{eq:hamiltonian_long_range_kitaev} remains quadratic and admits an exact single-particle description for any $\alpha$.
	
	\subsection{Bogoliubov-de Gennes formulation}
	\label{subsec:model_bdg}
	
	We rewrite Eq.~\eqref{eq:hamiltonian_long_range_kitaev} in Bogoliubov-de Gennes (BdG) form.
	We introduce the Nambu spinor (a column vector of dimension $2N$) \cite{Altland1997Jan}
	\begin{equation}
		\Psi=
		\begin{pmatrix}
			c_1\\ \vdots\\ c_N\\ c_1^\dagger\\ \vdots\\ c_N^\dagger
		\end{pmatrix},
	\end{equation}
	so that
	\begin{equation}
		H_{\mathrm{LRK}}=\frac{1}{2}\Psi^\dagger H_{\mathrm{BdG}}\Psi + \frac{1}{2}\mathrm{Tr}(K).
		\label{eq:bdg_quadratic_form}
	\end{equation}
	The trace term simply shifts the total energy and is dropped without loss of generality. The $2N \times 2N$ BdG Hamiltonian has the block structure
	\begin{equation}
		H_{\mathrm{BdG}}=
		\begin{pmatrix}
			K & \Delta\\
			-\Delta^\ast & -K^T
		\end{pmatrix},
		\label{eq:bdg_blocks}
	\end{equation}
	where $K$ is the single-particle hopping plus on site matrix and $\Delta$ is the pairing matrix (with $\Delta^\ast$ being the complex conjugate).
	For Eq.~\eqref{eq:hamiltonian_long_range_kitaev}, these turn out to be
	\begin{subequations}
		\label{eq:K_Delta_def}
		\begin{equation}
			K_{ij}=
			\begin{cases}
				2 \cos \theta, & i=j,\\[4pt]
				\displaystyle \frac{\sin\theta}{|i-j|^\alpha}, & i\neq j,
			\end{cases}
			\label{eq:K_def}
		\end{equation}
		\begin{equation}
			\Delta_{ij}=
			\begin{cases}
				\displaystyle -\frac{(1+\epsilon)\sin\theta}{|i-j|^\alpha}, & i<j,\\[6pt]
				\displaystyle +\frac{(1+\epsilon)\sin\theta}{|i-j|^\alpha}, & i>j,\\[6pt]
				\hfil 0,\hfil & i=j.
			\end{cases}
			\label{eq:Delta_def}
		\end{equation}
	\end{subequations}
	Thus $K$ is real and symmetric while $\Delta$ is real and antisymmetric, as required by fermionic statistics.
	The BdG matrix obeys the intrinsic particle-hole constraint
	\begin{equation}
		\tau_x\, H_{\mathrm{BdG}}^T\, \tau_x = - H_{\mathrm{BdG}},
		\label{eq:bdg_phs}
	\end{equation}
	where $\tau_x=\begin{pmatrix}0&\mathds{1}_N\\ \mathds{1}_N&0\end{pmatrix}$ is the Pauli matrix acting in Nambu (particle-hole) space and exchanges the particle and hole blocks of $H_{\mathrm{BdG}}$.
	This constraint implies that BdG eigenvalues occur in $\pm E$ pairs.
	
	The Nambu representation doubles the single-particle description by treating particles and holes on equal footing; Eq.~\eqref{eq:bdg_phs} encodes the resulting redundancy, so the physical spectrum is not doubled.

	\subsection{Majorana representation}
	\label{subsec:model_majorana}
	
	We define Majorana operators $\gamma_\mu$ ($\mu=1,\dots,2N$) sitewise by
	\begin{equation}
		\gamma_{2j-1} = \frac{c_j + c_j^\dagger}{\sqrt{2}},
		\qquad
		\gamma_{2j} = \frac{c_j^\dagger - c_j}{i\sqrt{2}}.
		\label{eq:majorana_def}
	\end{equation}
	They satisfy $\gamma_\mu^\dagger=\gamma_\mu$ and $\{\gamma_\mu,\gamma_\nu\}=\delta_{\mu\nu}$.
	Inverting Eq.~\eqref{eq:majorana_def} gives
	\begin{equation}
		c_j=\frac{\gamma_{2 j-1}-i \gamma_{2 j}}{\sqrt{2}},
		\qquad
		c_j^{\dagger}=\frac{\gamma_{2 j-1}+i \gamma_{2 j}}{\sqrt{2}}.
		\label{eq:inverse_majorana}
	\end{equation}
	In this basis, the Hamiltonian takes the standard quadratic Majorana form
	\begin{equation}
		H_{\mathrm{LRK}} = \frac{i}{2} \sum_{\mu,\nu=1}^{2N} \mathcal{H}_{M,\mu\nu} \gamma_\mu \gamma_\nu,
		\label{eq:H_majorana}
	\end{equation}
	where $\mathcal{H}_M$ is a real and antisymmetric $2N\times 2N$ matrix.

	\subsection{Liouvillian in the linear sector}
	\label{subsec:model_liouvillian}
	
	The Heisenberg equation of motion is $\frac{d\mathcal{O}}{dt}=i[H_{\mathrm{LRK}},\mathcal{O}]$.
	A crucial property of Eq.~\eqref{eq:H_majorana} is that the commutator algebra closes on operators linear in Majoranas (proved in Appendix~\ref{app:closure_linear_ops}).
	Specifically,
	\begin{equation}
		[H_{\mathrm{LRK}},\gamma_\ell]= i\sum_{m=1}^{2N}\mathcal{H}_{M,m\ell}\gamma_m,
		\label{eq:comm_H_gamma_main}
	\end{equation}
	with a derivation given in Appendix~\ref{app:majorana_commutator}.
	As a consequence, for any operator $\mathcal{O}(t)=\sum_{\ell=1}^{2N}u_\ell(t)\gamma_\ell$, the coefficients satisfy a closed linear equation
	\begin{equation}
		\frac{d u(t)}{dt}=-\mathcal{H}_M u(t),
		\qquad
		u(t)=e^{-\mathcal{H}_M t}u(0).
		\label{eq:eom_coeffs_main}
	\end{equation}
	
	Equivalently, the Liouvillian superoperator $\mathcal{L}[\cdot]\equiv[H_{\mathrm{LRK}},\cdot]$ acts on this linear subspace as multiplication by the Hermitian generator
	\begin{equation}
		L_{\mathrm{sp}} = i\mathcal{H}_M.
		\label{eq:single_particle_liouvillian}
	\end{equation}
	This single-particle representation is the starting point for the algorithmic adaptation of operator growth diagnostics via the operator Lanczos algorithm to single-particle picture, as developed in the next section.

\subsection{Exact Krylov structure of the short-range limit}
		\label{subsec:srk:anchor}
		
		The single-particle Liouvillian $L_\mathrm{sp} = i\mathcal{H}_M$ derived above admits a fully analytic treatment in the short-range limit $\alpha \to \infty$, $\epsilon = 0$.
		In this limit only nearest-neighbor couplings survive, 
		
		 Eq.~\eqref{eq:hamiltonian_long_range_kitaev} reduces to the standard Kitaev chain with hopping $t = \sin\theta$, pairing $\Delta = \sin\theta (1+\epsilon)$, and chemical potential $\mu = -2\cos\theta$ (explicit Hamiltonian in Appendix~\ref{app:srk_exact}, Eq.~\eqref{eq:H_srk_app}). For analytically tractable \(\epsilon=0\), we have the hopping-pairing balanced situation \(\Delta=t\). The corresponding \(H_M\) becomes a tridiagonal real antisymmetric matrix: acting on any Majorana mode \(\gamma_k\), the Liouvillian \(L_{\rm sp}\) couples it only to its two nearest neighbors \(k-1\) and \(k+1\), with alternating bond strengths \(\mu\) intra-site and \(2t\) inter-site. Equivalently, the balanced short-range Kitaev chain in its Majorana single-particle representation is exactly the SSH chain with intracell bond \(\mu\) and intercell bond \(2t\), and the Krylov recursion generated from \(\gamma_1\) simply preserves this alternating-bond matrix structure step by step. As a direct consequence, the Krylov algorithm seeded with the boundary Majorana \(\gamma_1\) traverses the Majorana chain sequentially, and the Lanczos off-diagonal coefficients simply read off those alternating bond strengths.
		The complete mathematical proof by strong induction is given in Appendix~\ref{app:srk_proof}; here we state the result.
		
		\paragraph{Exact thermodynamic-limit result.}
		For the short-range Kitaev chain at $\Delta = t$ (i.e.,~$\epsilon = 0$) where $\Delta$ and $t$ are the pairing and hopping terms respectively, with boundary seed $\gamma_1$, the Lanczos recursion produces, for every $n \geq 1$ and every system size $N \geq 1$:
		\begin{equation}
			b_{2n-1} = |\mu|, \qquad b_{2n} = 2|t|, \qquad a_n = 0.
			\label{eq:srk_exact_main}
		\end{equation}
		Here $\mu$ is the chemical potential. See Eq.~\eqref{eq:H_srk_app} in Appendix~\ref{app:srk_exact} where the relation with the notation used in Eq.~\eqref{eq:hamiltonian_long_range_kitaev} is established. There are no finite-size corrections, no transients, and no approximations.
		Equation~\eqref{eq:srk_exact_main} is exact for any $N$ and therefore remains exact in the thermodynamic limit as well.
		
		\paragraph{Exact topological order parameter.}
		Define the \textit{Krylov staggering parameter} as the logarithmic ratio of consecutive off-diagonal Lanczos coefficients:
		\begin{equation}
		\boxed{	\eta_n \equiv \ln\!\frac{b_{2n-1}}{b_{2n}} }.
			\label{eq:etadef_main}
		\end{equation}
		By Eq.~\eqref{eq:srk_exact_main}, this quantity is exactly constant in both $n$ and $N$:
		\begin{equation}
			\eta_n = \ln\!\frac{|\mu|}{2|t|} = \mathrm{const.}, \quad \forall\, n,\, N.
			\label{eq:srk_eta_exact_main}
		\end{equation}
		Its sign is an \emph{exact} topological indicator:
		\begin{equation}
			\mathrm{sgn}(\eta_n) =
			\begin{cases}
				-1 & |\mu| < 2|t| \quad \text{(topological)}, \\
				\phantom{-}0  & |\mu| = 2|t| \quad \text{(critical point)}, \\
				+1 & |\mu| > 2|t| \quad \text{(trivial)},
			\end{cases}
			\label{eq:srk_topo_criterion_main}
		\end{equation}
		in exact agreement with Kitaev's phase boundary~\cite{Kitaev2001,Alicea2012}. This is not a numerical observation or an asymptotic statement --- it is a theorem: the sign of $\eta$ encodes the topological phase of the short-range Kitaev chain exactly, derived purely from the Krylov recursion algebra.


	\subsection{Krylov--SSH correspondence and topological order parameter}
	\label{subsec:krylov_ssh_main}

It is known in the literature~\cite{Parker2019,Menzler2024Sep} that the Lanczos recursion maps operator dynamics onto a tight-binding problem on the Krylov chain, whose hopping amplitudes are the off-diagonal Lanczos coefficients \(b_n\). In generic systems this chain is semi-infinite and structurally featureless; here we prove something sharply stronger and exactly: for the short-range Kitaev chain at \(\Delta=t\), the balanced physical Kitaev chain in its Majorana single-particle representation is itself the paradigmatic SSH chain~\cite{Su1979,Asboth}, and the Krylov chain generated by the boundary Majorana seed is algebraically identical --- as a matrix, not merely in spirit --- to that same SSH chain. This identification is exact, holds for every system size $N$ (all the way up to the thermodynamic limit) and every recursion depth $n$, and is proved by strong induction in Appendix~\ref{app:srk_proof}--\ref{app:srk_ssh}. It implies that the topological phase of the Kitaev chain and the topological phase of the Krylov chain are the same object, read off from the same mathematical structure: when $|\mu| < 2|t|$, the Krylov chain is in the SSH topological phase, its left boundary site ($v_0 \leftrightarrow \gamma_1$) is undimerized, and this is precisely the condition for Majorana zero modes in the physical chain.
	
The exact flat Lanczos structure of Eq.~\eqref{eq:srk_exact_main} admits a direct geometric interpretation: the Majorana matrix $\mathcal{H}_M$ at $\Delta = t$ is a real antisymmetric tridiagonal matrix whose entries alternate between $|\mu|$ (intra-site, $\gamma_{2j-1}$--$\gamma_{2j}$) and $2|t|$ (inter-site, $\gamma_{2j}$--$\gamma_{2j+1}$). Under the identification $A_j \leftrightarrow \gamma_{2j-1}$, $B_j \leftrightarrow \gamma_{2j}$, $t_1 \leftrightarrow |\mu|$, $t_2 \leftrightarrow 2|t|$, the single-particle Liouvillian $L_\mathrm{sp} = i\mathcal{H}_M$ is algebraically identical to the SSH Hamiltonian~\cite{Su1979,Asboth} --- not an analogy, but an exact matrix equality confirmed to machine precision (Appendix~\ref{app:srk_ssh}). Because the induction proof of Appendix~\ref{app:srk_proof} shows that the Krylov recursion traverses the Majorana chain one site at a time (Eq.~\eqref{eq:chain_picture_app}), the Lanczos off-diagonal coefficients $b_n$ are simply the bond strengths read off sequentially, and the Krylov tridiagonal is structurally the same alternating chain. The three objects --- real-space Majorana chain, SSH Hamiltonian, and Krylov tridiagonal --- are therefore in exact one-to-one correspondence (see the dictionary in Appendix~\ref{app:srk_ssh}). The topological criterion follows immediately from SSH physics~\cite{Su1979,Asboth}: the SSH chain is topological whenever the intercell bond dominates ($t_2 > t_1$), which under the identification above is $2|t| > |\mu|$, i.e.\ $|\mu| < 2|t|$ --- exactly Kitaev's phase boundary~\cite{Kitaev2001}. The Krylov staggering parameter $\eta_n = \ln(|\mu|/2|t|)$, being the logarithmic ratio of the weak (intracell) to strong (intercell) bond, is therefore an exact, $n$-independent, size-independent topological order parameter of the short-range Kitaev chain at $\epsilon = 0$: its sign encodes the phase without any reference to bulk gap closure, winding numbers, or band topology (Appendix~\ref{app:srk_ssh}, Eq.~\eqref{eq:eta_ssh_app}). We emphasize the important caveat that this correspondence operates at the level of the single-particle Majorana matrix $L_\mathrm{sp}$, not the full many-body Hamiltonian; the Kitaev chain is a $p$-wave superconductor while the SSH chain is an insulator, and they differ in symmetry class and many-body structure. The precise statement --- and its structural origin in the cancellation of all next-nearest-neighbor bonds at $\Delta = t$ --- is detailed in Appendix~\ref{app:srk_ssh}.

\subsection{Motivation for the long-range study.}
Equation~\eqref{eq:srk_topo_criterion_main} establishes $\eta_n$ as an analytically grounded, exactly proven topological diagnostic in the tractable short-range balanced limit: there, $\eta_n$ is \emph{constant} in $n$ and its sign alone suffices to classify the phase. When long-range power-law couplings ($\alpha < \infty$) or a nonzero pairing imbalance ($\epsilon \neq 0$) are introduced, this flatness is broken --- $\eta_n$ acquires a nontrivial dependence on recursion depth $n$ --- and a constant-sign criterion no longer directly applies. The diagnostic must therefore be recast. The key insight, which the exact short-range result makes precise, is that each value of $\eta_n$ encodes the local Krylov SSH chain-type bonding character at recursion depth $n$: $\eta_n < 0$ means the Krylov chain is locally in the topologically strong-bond regime at step $n$, and $\eta_n > 0$ means it is locally in the trivially weak-bond regime. A \emph{sign flip} of $\eta_n$ as a function of $n$ therefore signals that the Krylov recursion is traversing a crossover between these two regimes --- the system's low-energy structure is being probed at both topological and trivial energy scales as the recursion depth increases, a hallmark of a system in or near the topological phase. Conversely, when $\eta_n$ retains a fixed positive sign for all $n$, the Krylov chain remains uniformly in the trivial-bond regime throughout the recursion, signaling a bulk-dominated trivial phase. The remainder of this paper tests this hypothesis across the full $(\theta, \alpha)$ phase diagram of the long-range Kitaev chain at $\epsilon \neq 0$, and establishes empirically --- motivated and anchored by the exact short-range proof --- that the presence or absence of sign flips in $\eta_n$ as a function of $n$ provides a robust, operationally sharp diagnostic of the edge-versus-bulk character of the excitation gap, without requiring any reference to winding numbers, band topology, or bulk gap closure.

	\section{Single-Particle Lanczos Algorithm}
	\label{sec:algorithm}
	
	This section derives and develops the numerical construction of the Lanczos algorithm and the associated Lanczos coefficients for the long-range Kitaev chain by exploiting the exact closure of Heisenberg dynamics on operators linear in Majorana modes.
	Accordingly, all Krylov dynamics reported here take place in the $2N$-dimensional coefficient space associated with operators linear in Majoranas.

	The exact short-range result derived in Sec.~\ref{subsec:srk:anchor} provides a stringent benchmark for this construction. In the balanced short-range limit $\Delta=t$, with seed $\gamma_1$, the algorithm must reproduce $b_{2n-1}=|\mu|$, $b_{2n}=2|t|$, and $a_n=0$ for every recursion depth and every system size, up to machine precision (for a detailed proof, see Appendix~\ref{app:srk_exact}). The same construction is then used without modification in the long-range regime, where no closed-form expression is available and the Lanczos data become the central diagnostic.

	\subsection{Krylov subspace from the Liouvillian}
	\label{subsec:krylov_definition}
	
	Let $\mathcal{L}[\cdot]\equiv[H_{\mathrm{LRK}},\cdot]$ denote the Liouvillian superoperator.
	For an initial operator $\mathcal{O}(0)$, the associated Krylov subspace is
	\begin{equation}
		\mathbb{K}(\mathcal{O}(0))
		=\mathrm{span}\left\{
		\mathcal{O}(0),\,
		\mathcal{L}\mathcal{O}(0),\,
		\mathcal{L}^2\mathcal{O}(0),\,
		\ldots
		\right\}.
		\label{eq:krylov_span}
	\end{equation}
	The Lanczos algorithm constructs an orthonormal basis of $\mathbb{K}(\mathcal{O}(0))$ in which $\mathcal{L}$ is represented by a real symmetric tridiagonal matrix.
	
	We restrict to the linear Majorana sector,
	\begin{equation}
		\mathcal{O}=\sum_{\mu=1}^{2N} u_\mu\,\gamma_\mu,
		\qquad
		u\in\mathbb{C}^{2N}.
		\label{eq:O_linear_majorana_alg}
	\end{equation}
	Using Eq.~\eqref{eq:comm_H_gamma_main}, the Liouvillian action reduces to multiplication by a $2N\times 2N$ matrix on the coefficient vector,
	\begin{equation}
		\mathcal{L}\mathcal{O} \;\widehat{=}\; L_{\mathrm{sp}}\,u,
		\qquad
		L_{\mathrm{sp}} \equiv i\mathcal{H}_M,
		\label{eq:Lsp_def}
	\end{equation}
	where $\widehat{=}$ denotes the identification induced by Eq.~\eqref{eq:O_linear_majorana_alg}.
	Since $\mathcal{H}_M$ is real and antisymmetric, $L_{\mathrm{sp}}$ is Hermitian.
	Consequently, the entire Krylov construction can be performed in the single-particle space $\mathbb{C}^{2N}$ without explicit many-body operators.

	\subsection{Seed operator and inner product}
	\label{subsec:seed_inner_product}
	
	To probe local operator growth we seed the Krylov recursion with a single boundary Majorana operator,
	\begin{equation}
		\mathcal{O}(0)=\gamma_1,
		\label{eq:seed_operator_gamma1}
	\end{equation}
	which is Hermitian and localized at the left edge of the chain.
	In the coefficient representation~\eqref{eq:O_linear_majorana_alg}, this corresponds to the unit vector
	\begin{equation}
		v_0 \equiv u(0)=(1,0,\ldots,0)^T,
		\qquad
		\|v_0\|=1.
		\label{eq:v0_gamma1}
	\end{equation}
	
	To measure the overlap between operators in the Lanczos recursion, we need an inner product.
	For operators linear in Majoranas, the natural choice is the infinite-temperature Hilbert--Schmidt product
	\begin{equation}
		\langle \mathcal{A},\mathcal{B}\rangle_{\mathrm{HS}}
		= \frac{1}{2^N}\,\mathrm{Tr}(\mathcal{A}^\dagger\mathcal{B}).
	\end{equation}
	In the coefficient space $\mathbb{C}^{2N}$, this reduces to the Euclidean inner product up to an overall factor
	\begin{equation}
		\langle v,w\rangle \equiv v^\dagger w,
		\label{eq:euclidean_inner_product}
	\end{equation}
	where the proportionality is derived in Appendix~\ref{app:alg_inner_product}.
	
	Numerically, we drop the prefactor $2^{-N}$ to avoid exponentially small numbers at large $N$.
	Since the Lanczos algorithm depends only on orthonormalized directions and relative norms, multiplying the
	inner product by any constant leaves the Lanczos coefficients unchanged.

	\subsection{Lanczos recursion and tridiagonalization}
	\label{subsec:lanczos_recursion}
	
	Starting from $v_0$ in Eq.~\eqref{eq:v0_gamma1}, the Lanczos algorithm generates an orthonormal sequence $\{v_n\}_{n=0}^{\mathcal{K}-1}$ spanning the Krylov subspace $\mathbb{K}(\mathcal{O}(0))$.
	With the convention $v_{-1}\equiv 0$ and $b_0\equiv 0$, the three-term recurrence reads
	\begin{equation}
		L_{\mathrm{sp}} v_n = b_n v_{n-1} + a_n v_n + b_{n+1} v_{n+1},
		\label{eq:lanczos_three_term}
	\end{equation}
	where the Lanczos coefficients are
	\begin{align}
		a_n &= \langle v_n, L_{\mathrm{sp}} v_n\rangle \in \mathbb{R},
		\label{eq:an_def}\\
		w_n &= L_{\mathrm{sp}} v_n - a_n v_n - b_n v_{n-1},
		\label{eq:wn_def}\\
		b_{n+1} &= \|w_n\|,
		\qquad
		v_{n+1} =
		\begin{cases}
			w_n/b_{n+1}, & b_{n+1}>0,\\
			0, & b_{n+1}=0.
		\end{cases}
		\label{eq:bn_def}
	\end{align}
	The recursion terminates when $b_{n+1}$ falls below a numerical tolerance where we have implemented partial reorthogonalization.
	In practice we use $10^{-7}$, and the Krylov dimension is bounded by $ \mathcal{K} \le 2N$.
    Furthermore, to improve stability of the algorithm, we enforce symmetries that are shared by the Krylov vectors $v_n$ at every step, directly after $L_{sp}$ is applied to $v_n$.
    For example, a single Majorana initial seed, will always retain it's parity order, meaning that all $v_{2n-1}$ will have the same parity as $v_1$ while $v_{2n}$ will have opposite parity.
	
	\paragraph{Diagonal coefficients for a Majorana seed.}
	
	For any Hermitian seed operator, all diagonal Lanczos coefficients vanish identically (proved in Appendix~\ref{app:alg_an_zero}).
	Since $\mathcal{O}(0)=\gamma_1$ is Hermitian, we have
	\begin{equation}
		a_n=0\quad\text{for all }n.
		\label{eq:an_zero_statement}
	\end{equation}
	This mirrors the standard many-body operator Lanczos construction based on repeated commutators: for Hermitian dynamics and a Hermitian seed, the Krylov/Lanczos representation is purely off-diagonal.
	In our single-particle formulation, recovering $a_n=0$ therefore provides a stringent benchmark check.
	
	Equivalently, the Lanczos projection of $L_{\mathrm{sp}}$ onto the Krylov basis,
	\begin{equation}
		T \equiv V^\dagger L_{\mathrm{sp}} V,
		\label{eq:T_def_local}
	\end{equation}
	is a real symmetric tridiagonal matrix with vanishing diagonal, where $V$ denotes the $2N\times \mathcal{K}$ matrix whose columns are the Lanczos vectors $(v_0,\ldots,v_{\mathcal{K}-1})$.
	A general operator-space proof for seeds satisfying $\mathcal{O}(0)^\dagger=\pm\mathcal{O}(0)$ is given in Appendix~\ref{app:alg_an_zero}.
	In the present coefficient-space setting, the same conclusion follows from the antisymmetry of $\mathcal{H}_M$ and the fact that the Lanczos vectors generated from a real seed alternate between real and purely imaginary vectors, implying $v_n^\dagger(i\mathcal{H}_M)v_n=0$.

	\subsection{Numerical implementation}
	\label{subsec:numerics_krylov}
	
	For an open chain of $N$ fermionic sites, we construct the $2N\times 2N$ Majorana single-particle generator $\mathcal{H}_M$ appearing in Eq.~\eqref{eq:H_majorana}, and hence the Hermitian single-particle Liouvillian matrix $L_{\mathrm{sp}}=i\mathcal{H}_M$ acting on the $2N$-component coefficient vector $u$ of operators linear in Majoranas.
	We then run the Lanczos recursion to obtain $\{b_n\}$ and the Krylov dimension $\mathcal{K}\le 2N$, form the corresponding $\mathcal{K} \times \mathcal{K}$ tridiagonal matrix $T$, and evolve $\phi(t)$ by diagonalizing $T$.
	The recursion is terminated when the coefficient sequence becomes numerically unstable, and only Lanczos coefficients that satisfy multiple rigorous stability criteria are retained in the analysis; once any criterion is violated, that coefficient and all subsequent ones are excluded (see Appendix~\ref{app:alg_numerical_stability} for details).
	Additional seed operators used to test robustness are listed in Appendix~\ref{app:alt_seeds}.
	
	We recall Eq.~\eqref{eq:euclidean_inner_product} which is equivalent up to an overall constant factor to the infinite-temperature Hilbert-Schmidt product, $\langle \mathcal{A},\mathcal{B}\rangle_{\mathrm{HS}}=\frac{1}{2^N}\mathrm{Tr}\!\left(\mathcal{A}^\dagger\mathcal{B}\right).$
	In the numerics we omit the prefactor $2^{-N}$ to avoid exponentially small normalizations at large $N$.
	This rescaling does not affect the Lanczos coefficients (and hence any derived quantities), since the Lanczos procedure depends only on ratios fixed by orthonormalization and is invariant under an overall constant rescaling of the inner product.

	\section{Physics of Krylov Structure for Long-Range Pairing in Majorana Basis}
	\label{sec:physics of Krylov Structure for Long-Ranged Pairing in Majorana Basis}

		The preceding section established an exact analytic anchor for everything that follows: in the hopping-pairing-balanced short-range limit, the Krylov staggering parameter $\eta_n \equiv \ln(b_{2n-1}/b_{2n})$ is an exact topological order parameter, constant in $n$ and fixed in sign by the phase of the Kitaev chain itself (Sec.~\ref{subsec:srk:anchor} and Appendix~\ref{app:srk_exact}). The purpose of the present section is to show how this exact short-range result can be operationally generalized to the long-range, pairing-imbalanced problem, where $\eta_n$ is no longer constant but its sign structure continues to encode whether the lowest excitation is boundary-dominated or bulk-dominated.

	This section introduces two independent approaches to the edge-versus-bulk distinction: a \emph{static} analysis via BdG eigenmodes (Sec.~\ref{subsec:edge_vs_bulk_gap}) and a \emph{dynamical} diagnostic via operator Lanczos coefficients in the Krylov subspace (Sec.~\ref{subsec:krylov_majorana_physics}).
	The central result of this work (Sec.~\ref{sec:results}) is that these two approaches, constructed from entirely different physical principles, produce the same phase diagram.

	\subsection{Edge gap versus bulk gap}
	\label{subsec:edge_vs_bulk_gap}
	
	Open boundary conditions allow for low energy excitations that are spatially localized near the ends of the chain.
	In a finite system these boundary excitations can appear at energies parametrically below the extended bulk continuum, and it is therefore useful to distinguish an edge gap from the conventional bulk gap.
	
	We work with the $2N\times 2N$ Bogoliubov de Gennes Hamiltonian $H_{\mathrm{BdG}}$ introduced in Sec.~\ref{subsec:model_bdg} and assume it has been diagonalized for an open chain.
	Let $\{E_\nu\}_{\nu=1}^{2N}$ denote its eigenvalues and $\{\Phi_\nu\}$ the corresponding normalized eigenvectors, explicitly given by
	\begin{equation}
		\begin{aligned}
			H_{\mathrm{BdG}}\,\Phi_\nu =& E_\nu\,\Phi_\nu,\\
			\Phi_\nu=&
			\begin{pmatrix}
				u_\nu\\ v_\nu
			\end{pmatrix},
			\qquad
			u_\nu,v_\nu\in\mathbb{C}^{N},
			\qquad
			\|\Phi_\nu\|=1.
		\end{aligned}
	\end{equation}
	Particle hole symmetry implies that if $E_\nu$ is an eigenvalue then $-E_\nu$ is also an eigenvalue.
	We therefore restrict attention to the positive energies and sort them as
	\begin{equation}
		0\leq E_1\le E_2\le \cdots \le E_N.
	\end{equation}
	
	To decide whether a given BdG mode is boundary localized we introduce an edge weight.
	Fix a small integer $\ell_{\mathrm{edge}} = \lfloor\sqrt{N}\rfloor \ll N$ and define
	\begin{equation}
		\begin{aligned}
			W_\nu^{\mathrm{edge}}
			=&	\sum_{j=1}^{\ell_{\mathrm{edge}}}\Bigl(|u_{\nu,j}|^2+|v_{\nu,j}|^2\Bigr) \\
			&+
			\sum_{j=N-\ell_{\mathrm{edge}}+1}^{N}\Bigl(|u_{\nu,j}|^2+|v_{\nu,j}|^2\Bigr),
			\quad	0\le W_\nu^{\mathrm{edge}}\le 1.
		\end{aligned}
		\label{eq:edge_weight_def}
	\end{equation}
	We classify a mode as edge localized when $W_\nu^{\mathrm{edge}}>\omega_{\mathrm{edge}}$ for a fixed threshold
	$\omega_{\mathrm{edge}}$ and as bulk extended otherwise. Further discussions and physical meaning behind the operational parameters $\ell_{\rm edge}$ and $\omega_{\rm edge}$ are discussed in Sec.~\ref{subsec:phase diagrams}.

	With this classification and motivated by Ref.~\cite{Vodola2015Dec}, we define the edge and bulk gaps by
	\begin{equation}
		\Delta_{\mathrm{edge}}
		\equiv
		\min_{\nu\in\mathrm{edge}} E_\nu,
		\qquad
		\Delta_{\mathrm{bulk}}
		\equiv
		\min_{\nu\in\mathrm{bulk}} E_\nu.
		\label{eq:edge_bulk_gaps_def}
	\end{equation}
	By construction $\Delta_{\mathrm{edge}}$ measures the energy scale of the lowest boundary localized BdG excitation, whereas $\Delta_{\mathrm{bulk}}$ measures the onset of the extended bulk spectrum.
	
	We use the relative magnitude of these two scales to classify parameter regimes.
	We refer to a parameter point as belonging to the \emph{edge gap phase} when
	\begin{equation}
		\Delta_{\mathrm{edge}} < \Delta_{\mathrm{bulk}},
		\label{eq:edge_gap_phase}
	\end{equation}
	indicating that the lowest positive energy excitation is boundary localized.
	Conversely, we refer to the \emph{bulk gap phase} when
	\begin{equation}
		\Delta_{\mathrm{bulk}} \le \Delta_{\mathrm{edge}},
		\label{eq:bulk_gap_phase}
	\end{equation}
	indicating that the lowest excitation is extended in the bulk.
	This classification provides an operational diagnostic for whether low energy physics is dominated by boundary or bulk degrees of freedom \cite{Vodola2015Dec}. Numerical details and further discussions about $\omega_{\mathrm{edge}}$ and $\ell_{\mathrm{edge}}$ are provided in Section \ref{subsec:phase diagrams} and data for alternative seed are available in Appendix \ref{app:alt_seeds}.
	
	To summarize the construction: for each parameter point $(\alpha,\theta)$ we diagonalize $H_{\mathrm{BdG}}$ to obtain the complete set of eigenstates and their energies. Each eigenstate is then assigned an edge weight $W_\nu^{\mathrm{edge}}$ via Eq.~\eqref{eq:edge_weight_def}, which quantifies the fraction of its amplitude concentrated near the chain boundaries. States with $W_\nu^{\mathrm{edge}} > \omega_{\mathrm{edge}}$ are classified as edge-localized, while the remainder are classified as bulk-extended. From these two subsets we extract the minimum energy in each class, yielding $\Delta_{\mathrm{edge}}$ and $\Delta_{\mathrm{bulk}}$. The parameter point is then assigned to the edge gap phase or bulk gap phase according to which scale is smaller. This procedure follows the physical logic established in Ref.~\cite{Vodola2015Dec}, where the competition between boundary and bulk energy scales was identified as the key diagnostic for long-range topological phases, which is reproduced and validated by our methodology and construction.

	Having established the static BdG-based classification, we now introduce a completely independent dynamical diagnostic based on operator growth in Krylov space, and show later in Sec.~\ref{sec:results} that it yields the same phase diagram.

	\subsection{Krylov subspace and edge sensitivity}
	\label{subsec:krylov_majorana_physics}
	
	The Heisenberg dynamics of operators linear in Majoranas closes exactly in the single-particle coefficient space, with generator $L_{\mathrm{sp}} = i\mathcal{H}_M$, where $\mathcal{H}_M$ is real and antisymmetric.
	Consequently, $L_{\mathrm{sp}}$ is Hermitian.
	For a real boundary seed $v_0$ corresponding to $\mathcal{O}(0)=\gamma_1$, the Krylov vectors produced by the Lanczos algorithm alternate between real and purely imaginary vectors: $v_{2n}$ can be chosen real, while $v_{2n+1}$ can be chosen purely imaginary.
	This follows from $L_{\mathrm{sp}} = i\mathcal{H}_M$ with $\mathcal{H}_M$ real: applying $L_{\mathrm{sp}}$ to a real vector yields a purely imaginary vector, and applying $L_{\mathrm{sp}}$ again returns a real vector.
	This bipartite structure partitions Krylov space into two invariant subspaces under $L_{\mathrm{sp}}^2$:
	
	\begin{equation}
		\begin{aligned}
			\mathbb{K}_{\mathrm{even}} &= \operatorname{span}\{v_0,\,L_{\mathrm{sp}}^2 v_0,\,L_{\mathrm{sp}}^4 v_0,\dots\}, \\
			\mathbb{K}_{\mathrm{odd}} &= \operatorname{span}\{L_{\mathrm{sp}} v_0,\,L_{\mathrm{sp}}^3 v_0,\dots\}.
		\end{aligned}
	\end{equation}
	
	The Lanczos recursion on $L_{\mathrm{sp}}$ interleaves these two subspaces, producing a single tridiagonal chain with vanishing diagonal elements $a_n=0$ (see Appendix \ref{app:alg_an_zero}). The off-diagonal coefficients $\{b_n\}$ encode how these two subspaces couple at each step.

		To quantify the imbalance between the odd and even Krylov subsequences, we use the \emph{Krylov staggering parameter} $\eta_n$ defined in Eq.~\eqref{eq:etadef_main}, evaluated only within the numerically stable window and with $n\ge 2$ to suppress initialization effects. In the analytically solvable short-range balanced limit ($\epsilon=0$, $\alpha\to\infty$), $\eta_n$ is exactly constant and its sign is an exact topological order parameter (Sec.~\ref{subsec:srk:anchor}); the long-range problem studied here asks how that theorem deforms once long-range couplings ($\alpha<\infty$) and pairing imbalance ($\epsilon\neq 0$) render $\eta_n$ recursion-depth dependent. The central empirical finding established in Sec.~\ref{sec:results}, motivated directly by the exact short-range result, is that while the flatness of $\eta_n$ is lost, the \emph{sign structure} of $\eta_n$ across $n$ continues to distinguish regimes governed predominantly by edge-localized low-energy physics from those governed by bulk-extended excitations.

	The analysis window is restricted to the numerically stable portion of the recursion; in particular, we exclude all coefficients once $b_n \lesssim 10^{-7}$ or any other stability criterion is violated (see Appendix~\ref{app:alg_numerical_stability} for the complete set of criteria).
	The lower bound $n\geq 2$ excludes the first recursion step, where initialization effects can dominate.
	When $\eta_n \approx 0$, the odd and even subsequences are approximately equal; deviations from zero indicate asymmetry between the two invariant subspaces.
	The operational definition of sign changes and the choice of analysis window are detailed in Sec.~\ref{subsec:phase diagrams}.

	\paragraph{\texorpdfstring{Factors influencing $\eta_n$.}{Factors influencing staggering parameters}} Several mechanisms can break this symmetry:
	
	\begin{itemize}
		\item Pairing ($\epsilon \neq -1$): Breaks $U(1)$ charge conservation and mixes particle and hole sectors in $H_{\mathrm{BdG}}$, generically removing any structural equivalence between the restrictions of $L_{\mathrm{sp}}^2$ to $\mathbb{K}_{\mathrm{even}}$ and $\mathbb{K}_{\mathrm{odd}}$.
		
		\item Competition between energy scales: Even in the particle-conserving limit ($\epsilon = -1$), a nonzero $\eta_n$ can arise when the on-site energy $2\cos\theta$ competes with the kinetic scale $\sin\theta$, so that the Krylov recursion does not treat the two invariant subspaces on an equal footing.

		\item Long-range couplings ($\alpha \lesssim 1$): Power-law decaying matrix elements enhance hybridization between edge and bulk modes, causing boundary states to acquire algebraically decaying profiles. In the Krylov recursion, this extended hybridization can delay the emergence of clear even/odd splitting to larger $n$, whereas short-range models ($\alpha \gtrsim 2$) typically exhibit well-defined staggering at small Krylov depth.
	\end{itemize}

	Consequently, we interpret $\eta_n$ as an operational measure of how unevenly the Krylov recursion distributes operator weight between the odd and even subsequences, rather than as a quantity controlled by a single microscopic parameter.

	\paragraph{Boundary sensitivity and connection to edge physics.} A boundary seed $\gamma_1$ has support concentrated at the chain end. When the system hosts a low-energy edge-localized BdG mode, the seed couples preferentially to that mode, imprinting a characteristic pattern on the Lanczos coefficients $\{b_n\}$. This makes $\eta_n$ sensitive to whether the lowest excitation is boundary-localized or extended in the bulk.
	
	To quantify this sensitivity, we compare the sign structure of $\eta_n$ with the edge/bulk gap classification of Sec.~\ref{subsec:edge_vs_bulk_gap}. For the pairing-imbalanced models studied here ($\Delta \neq 0$), we observe a consistent empirical pattern:
	
	\begin{itemize}[leftmargin=*, itemsep=4pt]
		\item In the \emph{bulk gap phase} ($\Delta_{\mathrm{bulk}} \leq \Delta_{\mathrm{edge}}$), $\eta_n$ typically maintains a fixed sign (or shows no systematic sign changes) throughout the accessible recursion depth.
		
		\item In the \emph{edge gap phase} ($\Delta_{\mathrm{edge}} < \Delta_{\mathrm{bulk}}$), $\eta_n$ exhibits nonzero sign changes as $n$ increases. 
	\end{itemize}
	
	The diagnostic power of this pattern is clearest when the edge mode is well separated from the bulk continuum (e.g., when hopping and pairing strengths are comparable): the seed couples to a distinct low-energy feature at small $n$, yielding a sharp, numerically stable signature in $\eta_n$. When the edge and bulk gaps become comparable (e.g., when pairing dominates overwhelmingly over hopping), the qualitative correspondence between $\eta_n$ and the gap-based classification persists, but quantitative matching becomes numerically challenging, as the odd and even sub-chain responses nearly coincide within finite resolution. Nonetheless, $\eta_n$ remains qualitatively sensitive to boundary physics (see the discussion in Sec.~\ref{subsec:phase diagrams} and Appendix~\ref{app:alt_seeds} for analysis of alternative seeds); the underlying BdG eigenmode structure of these regimes is analyzed in Sec.~\ref{subsec:spectrum_impact_lr} and Appendix~\ref{app:spectra_additional_epsilon}.

	\paragraph{Quantifying sign changes.} To count sign changes robustly, we introduce a tolerance $\eta_{\mathrm{tol}} > 0$ and define a discrete sign variable
	\begin{equation}
		s_n =
		\begin{cases}
			+1, & \eta_n > \eta_{\mathrm{tol}},\\
			-1, & \eta_n < -\eta_{\mathrm{tol}},\\
			0, & |\eta_n| \le \eta_{\mathrm{tol}}.
		\end{cases}
		\label{eq:eta_tol_introduced}
	\end{equation}
	Consider the window $n_{\min} \le n \le n_{\max}$ chosen to avoid any finite-size boundary effects. In practice, we take $n_{\min}=2$ to exclude the first few steps, and choose $n_{\max}$ as the largest index up to which the Lanczos coefficients remain numerically stable (see Appendix~\ref{app:alg_numerical_stability}).
	Implementation details are provided in Section \ref{subsec:phase diagrams}.
	Let $\{s_{n_1}, s_{n_2}, \dots, s_{n_M}\}$ be the subsequence of \emph{nonzero} signs extracted from $\{s_n\}_{n_{\min} \le n \le n_{\max}}$ and ordered by increasing $n$ (so $n_1 < n_2 < \dots < n_M$). The \emph{crossing count}
	\begin{equation}
		N_{\mathrm{cross}} = \sum_{k=1}^{M-1} \Theta \bigl(-s_{n_k} s_{n_{k+1}}\bigr),
		\label{eq:Ncross_def}
	\end{equation}
	tallies the number of sign flips within this filtered subsequence, chosen within the numerically stable part of the recursion (see App.~\ref{app:algorithm_details} for the discussion on numerical stabitility). Here $\Theta(x)$ is the Heaviside step function ($\Theta(x)=1$ for $x>0$, $0$ otherwise), so each term contributes $1$ precisely when $s_{n_k}$ and $s_{n_{k+1}}$ have opposite signs.
	
	In practice, $N_{\mathrm{cross}}$ depends strongly on the seed location: boundary seeds bias the recursion toward one chain end, while bulk seeds couple comparably to both ends and can produce multiple sign flips at parametrically separated depths. The mapping between $\eta_n$ and edge/bulk gap classification is therefore quantitatively most reliable when using a boundary seed; data for alternative seed are provided in Appendix \ref{app:alt_seeds}.

	In summary, the Krylov staggering parameter $\eta_n$ provides a \emph{dynamical probe of the (static) boundary versus bulk character of low-energy excitations} that complements static gap measurements.
	Its sign structure correlates with the edge/bulk character of the lowest excitation, offering a dynamical, Krylov subspace diagnostic of edge dominance even when the bulk gap does not close.

	\begin{figure*}[t]
		\centering
		
		\begin{overpic}{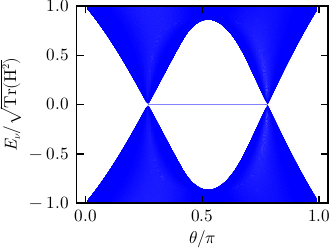}
			\put(32,66){\color{white}(a) $\alpha=3$}
		\end{overpic}
		{\phantomsubcaption\label{fig:spectrum_alpha30_eps_minus02}}
		\begin{overpic}{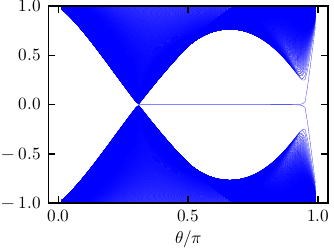}
			\put(27,66){\color{white}(b) $\alpha=1$}
		\end{overpic}
		{\phantomsubcaption\label{fig:spectrum_alpha10_eps_minus02}}
		\begin{overpic}{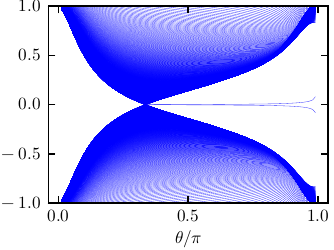}
			\put(27,66){\color{white}(c) $\alpha=1/3$}
		\end{overpic}
		{\phantomsubcaption\label{fig:spectrum_alpha013_eps_minus02}}
		\caption{
			BdG spectrum $E_\nu/\sqrt{\operatorname{Tr}(H^2)}$ versus $\theta/\pi$ at $\epsilon=-0.2$ for three long-range exponents $\alpha$ ($N=1000$, open boundaries). 
			\subref{fig:spectrum_alpha30_eps_minus02} Short-range-like behavior ($\alpha=3$) displays a gapless region centered at $\theta/\pi \approx 0.5$ with sparse spectral density near $E=0$ elsewhere. 
			\subref{fig:spectrum_alpha10_eps_minus02} Intermediate regime ($\alpha = 1$) retains a similar gapless range with modified spectral density. 
			\subref{fig:spectrum_alpha013_eps_minus02} Strong long-range limit ($\alpha=1/3$) lifts the degeneracy throughout the $\theta/\pi$ interval except for a small near-degenerate regime, with markedly denser spectral density near $E=0$ throughout.
		}
		\label{fig:spectrum_eps_minus02}
	\end{figure*}
	
	\section{Results}
	\label{sec:results}
	\subsection{Spectrum and the role of long-range pairing}
	\label{subsec:spectrum_impact_lr}
	
	\Cref{fig:spectrum_eps_minus02} shows the Bogoliubov-de~Gennes (BdG) eigenenergies $E_\nu$,
	normalized by $\sqrt{\mathrm{Tr}(H^2)}$ to facilitate comparison across different $\alpha$ values, as a function of the twist parameter $\theta/\pi$ at fixed
	$\epsilon=-0.2$ for three long-range exponents $\alpha$ (open boundaries, $N=1000$).
	The particle-hole symmetry $\tau_x H_{\mathrm{BdG}}^T \tau_x = -H_{\mathrm{BdG}}$ (see Eq.~\eqref{eq:bdg_phs}) ensures that the spectrum is symmetric under $E\mapsto -E$.
	
	\paragraph{\texorpdfstring{Short-range-like spectrum ($\alpha=3$, \cref{fig:spectrum_alpha30_eps_minus02}).}{Short-range-like spectrum}}
	The spectrum exhibits a gapless region over a finite range of $\theta/\pi$ roughly centered at $\theta/\pi \approx 0.5$, consistent with the short-range Kitaev chain~\cite{Kitaev2001}.
	Outside this gapless regime, the system is gapped and the spectral density near $E=0$ remains comparatively sparse, reflecting exponentially localized boundary modes and a band structure with weak hybridization across the chain length scale.
	
	\paragraph{\texorpdfstring{Intermediate exponent ($\alpha=1$, \cref{fig:spectrum_alpha10_eps_minus02}).}{Intermediate exponent}}
	The system remains gapless over a $\theta/\pi$ range similar to that of the short-range case, with modifications to the low-energy spectral structure.
	This marks a crossover between short-range-like and long-range-dominated behavior, where power-law couplings begin to alter the eigenmode structure.

	\paragraph{\texorpdfstring{Strong long-range regime ($\alpha=1/3$, \cref{fig:spectrum_alpha013_eps_minus02}).}{Strong long-range regime.}}
	Long-range pairing lifts the degeneracy throughout the $\theta/\pi$ interval, with the exception of a small residual near-degeneracy regime as the system is made further long-ranged, the leftmost point of the gapless regime observed at larger $\alpha$.
	Elsewhere, regions that were gapless for $\alpha \ge 1$ become gapped, and the spectral density near $E=0$ increases markedly across the entire $\theta/\pi$ range.
	This behavior is consistent with the edge-mode mass acquisition mechanism documented in Refs.~\cite{Vodola2014Oct, Vodola2015Dec} for $\alpha < 1$, where purely algebraic spatial decay of correlations replaces the hybrid exponential-algebraic decay.

	\paragraph*{Implications for Krylov dynamics.} Comparing these spectra with the edge-weight classification of Sec.~\ref{subsec:edge_vs_bulk_gap} (results shown later in Sec.~\ref{subsec:phase diagrams}) reveals how the edge-dominated versus bulk-dominated regimes shift across the $(\alpha,\theta)$ phase diagram.
	For short-range behavior ($\alpha=3$, Fig.~\ref{fig:spectrum_alpha30_eps_minus02}), the edge gap phase---where the lowest positive BdG mode carries significant edge weight---is confined to the gapless region centered roughly around $\theta/\pi \sim 0.5$.
	Outside this window, where the system is gapped, the lowest excitation is bulk extended and the system enters the bulk gap phase.
	As $\alpha$ decreases toward the strong long-range limit ($\alpha=1/3$, Fig.~\ref{fig:spectrum_alpha013_eps_minus02}), the edge gap phase extends over a much broader range of $\theta$, so that edge-localized modes dominate the low-energy physics across a larger portion of the phase diagram than in the short-range case.
	The central result of this work is that this expansion of the edge-dominated regime in parameter space, extracted from the BdG analysis, is precisely reproduced by the \textit{independent} Krylov-based crossing-count diagnostic for the staggering parameter $\eta_n$ (as shown in Sec.~\ref{sec:results}).

	Spectra at $\epsilon=1$ and $\epsilon=10$ (Appendix~\ref{app:spectra_additional_epsilon}) confirm these trends persist across pairing regimes, without altering the fundamental $\alpha$-driven eigenmode reorganization. Therefore, we focus henceforth on $\epsilon=-0.2$, where moderate density near $E=0$ facilitates numerical resolution of the odd- and even-parity Krylov subspaces while preserving the essential long-range physics.

	\begin{figure*}[htbp]
		\centering
		\begin{overpic}{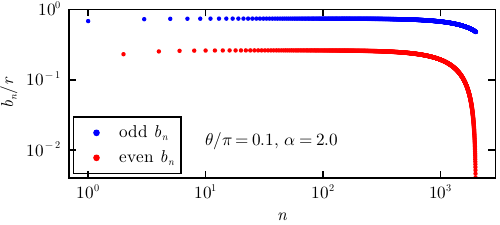}
			\put(43,22){(a) bulk gap}
		\end{overpic}
		\begin{overpic}{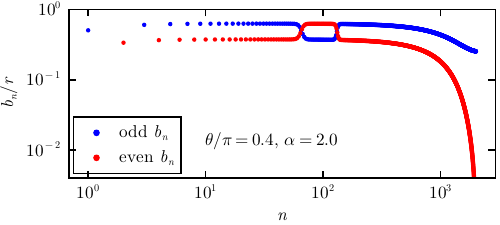}
			\put(43,22){(c) edge gap}
		\end{overpic}
		\vspace{2mm}
		
		\begin{overpic}{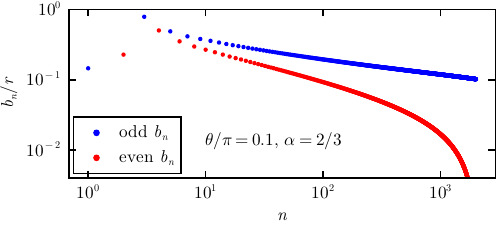}
			\put(43,22){(b) bulk gap}
		\end{overpic}
		\begin{overpic}{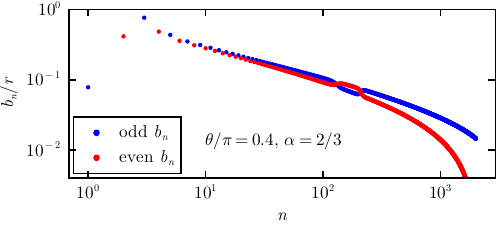}
			\put(43,22){(d) edge gap}
		\end{overpic}
		{\phantomsubcaption\label{fig:lanczos_coefficients_shortrange_trivial}}
		{\phantomsubcaption\label{fig:lanczos_coefficients_longrange_trivial}}
		{\phantomsubcaption\label{fig:lanczos_coefficients_shortrange_edge}}
		{\phantomsubcaption\label{fig:lanczos_coefficients_longrange_edge}}

		\caption{%
			Lanczos coefficients $\{b_n\}$ for the long-range Kitaev chain at $\epsilon=-0.2$ with open boundaries and Hermitian boundary seed $\gamma_1$ ($N=1000$).
			Each panel shows the two interleaved subsequences (odd and even steps of the recursion), whose relative ordering determines the sign of the staggering parameter $\eta_n=\ln(b_{2n-1}/b_{2n})$ and hence the crossing count $N_{\mathrm{cross}}$ (Eq.~\eqref{eq:Ncross_def}).
			Panels \subref{fig:lanczos_coefficients_shortrange_trivial}, \subref{fig:lanczos_coefficients_longrange_trivial} show parameter points in the bulk-gap regime (discussed later in the context of Fig.~\ref{fig:lrk_edge_bulk_phase_diagram}) and exhibit no interchange of the two subsequences (consistent with $N_{\mathrm{cross}}=0$), while panels \subref{fig:lanczos_coefficients_shortrange_edge}, \subref{fig:lanczos_coefficients_longrange_edge} lie in the edge-gap regime and show a clear interchanges (consistent with $N_{\mathrm{cross}}\ge 1$). To maintain numerical rigor, the Lanczos recursion is terminated when $b_n \lesssim 10^{-7}$, and all subsequent coefficients are excluded from the analysis. Consequently, the total number of Lanczos coefficients varies across parameter points (see Appendix~\ref{app:alg_numerical_stability} for stability criteria).
		}
		\label{fig:lanczos_coeffs_maj1}
	\end{figure*}

	\subsection{Lanczos Coefficients}
	\label{subsec:Lanczos Coefficients}

		Before turning to the long-range data, we recall the exact short-range anchor established in Sec.~\ref{subsec:srk:anchor}: for the balanced short-range chain ($\alpha\to\infty$, $\epsilon=0$), the Lanczos coefficients are exactly flat in alternating pairs, $b_{2n-1}=|\mu|$ and $b_{2n}=2|t|$ for every $n$ and every system size $N$ (Eq.~\eqref{eq:srk_exact_main}). Equivalently, the staggering parameter $\eta_n=\ln(b_{2n-1}/b_{2n})$ is exactly constant, and its sign is an exact topological order parameter of the short-range Kitaev chain (Eq.~\eqref{eq:srk_topo_criterion_main}). The long-range data below show how this flat structure deforms once power-law couplings are introduced, while preserving the central physical lesson of the short-range theorem: the sign structure of $\eta_n$, and hence the crossing count $N_{\mathrm{cross}}$ defined in Eq.~\eqref{eq:Ncross_def}, continues to track whether the lowest excitation is edge-dominated or bulk-dominated across the full $(\theta,\alpha)$ phase diagram.

	We now present representative Lanczos-coefficient data that underlie the crossing-count diagnostic and connection to edge-versus-bulk gap physics. 
	For the Hermitian boundary seed $\gamma_1$, all diagonal Lanczos coefficients vanish, $a_n=0$ (Appendix~\ref{app:alg_an_zero}), so the Krylov subspace representation is purely off-diagonal and is fully characterized by the sequence $\{b_n\}$. 
	In this setting it is natural to view the recursion as two interleaved ``sub-chains'' (odd and even steps), whose relative ordering is quantified by the staggering parameter $\eta_n=\ln(b_{2n-1}/b_{2n})$ ($n\geq 2$), as introduced in Sec.~\ref{subsec:srk:anchor} and further discussed in Sec.~\ref{subsec:krylov_majorana_physics}.
	All data presented satisfy rigorous numerical stability criteria; in particular, we exclude all Lanczos coefficients once $b_n \lesssim 10^{-7}$ or any other stability check fails (see Appendix~\ref{app:alg_numerical_stability} for details).

\begin{figure}[htbp]
		\centering
		\includegraphics[]{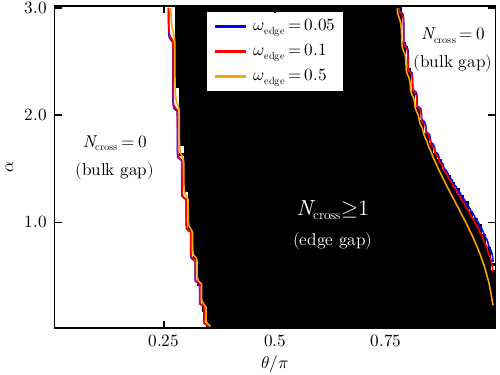} 
		\caption{
			Phase diagram for the long-range Kitaev chain at $\epsilon=-0.2$ with open boundaries ($N=1000$) and boundary seed $\gamma_1$, generating $2000$ Lanczos coefficients.
			As discussed in Sec.~\ref{subsec:krylov_majorana_physics}, the black region indicates parameters where the Krylov staggering parameter $\eta_n=\ln(b_{2n-1}/b_{2n})$ exhibits nonzero robust sign changes ($N_{\mathrm{cross}}\ge 1$), while the white region corresponds to $N_{\mathrm{cross}}=0$.
			As discussed in Sec.~\ref{subsec:edge_vs_bulk_gap}, solid curves show the edge-bulk gap boundary $\Delta_{\mathrm{edge}}=\Delta_{\mathrm{bulk}}$ extracted from the BdG spectrum using the edge-weight criterion for three choices of threshold, $\omega_{\mathrm{edge}}=0.05,0.1,0.5$ (with $\ell_{\mathrm{edge}}=\sqrt{N}$).
			The central numerical result is the near coincidence of these independently constructed boundaries: the Krylov-based and gap-based boundaries coincide within the numerical resolution of the $99\times 99$ grid in $(\alpha,\theta)$ with $\alpha\in(0,3]$ and $\theta\in(0,\pi)$.
			The phase boundary is robust across different $\omega_{\mathrm{edge}}$ values; discrepancies arise from grid resolution and finite-size effects, the latter being most pronounced in the strong long-range regime ($\alpha\lesssim 1$) at larger $\theta$ (see main text for further discussion).
		}
        \label{fig:lrk_edge_bulk_phase_diagram}
	\end{figure}

	Figure~\ref{fig:lanczos_coeffs_maj1} shows $\{b_n\}$ at four parameter points chosen to span short-range-like and long-range regimes as well as bulk-gap and edge-gap behavior (as classified in Fig.~\ref{fig:lrk_edge_bulk_phase_diagram}).
	In the bulk-gap cases, panels~\subref{fig:lanczos_coefficients_shortrange_trivial} ($\alpha=2,\ \theta/\pi=0.1$) and \subref{fig:lanczos_coefficients_longrange_trivial} ($\alpha=2/3,\ \theta/\pi=0.1$), the two sub-chains do not interchange their ordering over the stable recursion window, and correspondingly $\eta_n$ does not undergo a sign flip (yielding $N_{\mathrm{cross}}=0$).
	By contrast, in the edge-gap cases, panels~\subref{fig:lanczos_coefficients_shortrange_edge} ($\alpha=2,\ \theta/\pi=0.4$) and \subref{fig:lanczos_coefficients_longrange_edge} ($\alpha=2/3,\ \theta/\pi=0.4$), the two sub-chains clearly interchange, producing nonzero sign changes in $\eta_n$ and hence $N_{\mathrm{cross}}\ge 1$.
	The same qualitative pattern appears in both the short-range-like and long-range choices of $\alpha$, indicating that it is the edge-versus-bulk control of the lowest excitation scale---not the mere presence of long-range couplings---that governs whether crossings occur.
	
	Having demonstrated this behavior at representative parameter points, we now compute the crossing-count phase diagram $N_{\mathrm{cross}}(\alpha,\theta)$ (Eq.~\eqref{eq:Ncross_def}) systematically across the full parameter space on a uniform grid (as discussed in Sec.~\ref{subsec:krylov_majorana_physics}) and compare it with the BdG-based edge-bulk gap phase diagram (as discussed in Sec.~\ref{subsec:edge_vs_bulk_gap}), finding quantitative agreement within our numerical resolution.

	\subsection{Edge-Bulk Gap vs. Krylov Staggering Phase Diagrams}
	\label{subsec:phase diagrams}
	
	In this subsection, we compare a \emph{static} classification of low-energy excitations, based on whether the smallest positive BdG mode is boundary localized or bulk extended, with a \emph{dynamical} classification extracted from single-particle operator Lanczos algorithm.
	Concretely, we overlay the edge-bulk gap phase diagram (detailed in Sec.~\ref{subsec:edge_vs_bulk_gap}) with the crossing count obtained from the Krylov staggering parameter (detailed in Sec.~\ref{subsec:krylov_majorana_physics}). This is shown in Figure~\ref{fig:lrk_edge_bulk_phase_diagram}. 
	
	Unless stated otherwise, the phase diagrams are computed for an open chain with $N=1000$ and a boundary seed operator, and are evaluated on a uniform $99 \times 99$ grid in $(\alpha,\theta)$ where $\alpha \in (0, 3]$ and $\theta \in (0, \pi)$. Boundary-localized seeds provide the sharpest quantitative agreement with the BdG-derived edge-bulk boundary (up to grid resolution and finite-size effects), whereas bulk seeds couple comparably to both edges and typically yield less sharp matching. Data for additional seeds are provided in Appendix~\ref{app:alt_seeds}. 
			Finite-size robustness of the extracted boundary is demonstrated in two complementary ways in Sec.~\ref{app:finite_size_scaling}.

	Before we discuss the results, we provide a brief discussion about the operational parameters used to define the edge weight and to count sign changes in $\eta_n$.

	\paragraph{Operational edge-weight cutoff.}
	To determine whether a given Bogoliubov-de Gennes (BdG) eigenmode is boundary localized in a finite open chain, we quantify its boundary support by the edge weight as defined in Eq.~\eqref{eq:edge_weight_def}. There $\ell_{\mathrm{edge}}\ll N$ is a fixed boundary-window size.
	We then \emph{classify} mode $\nu$ as edge localized if
	\begin{equation}
		W_{\nu}^{\mathrm{edge}}>\omega_{\mathrm{edge}},
	\end{equation}
	for a threshold $\omega_{\mathrm{edge}}\in(0,1)$, and as bulk extended otherwise.
	Importantly, $\ell_{\mathrm{edge}}$ and $\omega_{\mathrm{edge}}$ are \emph{operational} parameters rather than universal constants: they define a practical partition of the finite-size spectrum into modes with predominantly boundary support versus modes that are spatially extended.
	
	\paragraph{Robustness and physical meaning.}
	The physical content of this criterion lies in its stability under controlled variations of $(\ell_{\mathrm{edge}},\omega_{\mathrm{edge}})$ and system size $N$.
	For bulk-extended modes, normalization implies that the typical weight contained in two fixed boundary windows scales as
	$W_{\nu}^{\mathrm{edge}}\sim 2\ell_{\mathrm{edge}}/N$, hence $W_{\nu}^{\mathrm{edge}}\to 0$ as $N\to\infty$ at fixed $\ell_{\mathrm{edge}}$.
	By contrast, for a boundary-localized mode whose localization length remains $O(1)$ as $N\to\infty$, the edge weight does not scale as $\ell_{\mathrm{edge}}/N$; instead $W_{\nu}^{\mathrm{edge}}$ stays finite (i.e., does not vanish with $N$) for fixed $\ell_{\mathrm{edge}}$.
	Consequently, whenever the finite-size spectrum exhibits a clear separation between the distributions of $W_{\nu}^{\mathrm{edge}}$ for edge-like and bulk-like states, the overall phase structure is robust to variations in $\omega_{\mathrm{edge}}$.
	We fix $\ell_{\mathrm{edge}} = \lfloor\sqrt{N}\rfloor$ throughout and test this stability by varying $\omega_{\mathrm{edge}}$ from $0.05$ to $0.5$. As demonstrated in Fig.~\ref{fig:lrk_edge_bulk_phase_diagram}, the qualitative distinction between edge-gap and bulk-gap regions persists across this range, while the precision of the extracted boundary improves as the threshold is lowered.
	The residual sensitivity to $\omega_{\mathrm{edge}}$ is most pronounced in the strong long-range regime ($\alpha < 1$) at large $\theta/\pi$, consistent with finite-size effects where algebraically decaying eigenmodes extend over a larger fraction of the chain.

	\paragraph{\texorpdfstring{Numerical resolution for $N_{\mathrm{cross}}$.}{Numerical resolution of number of crossings}}
	The crossing count $N_{\mathrm{cross}}$ in Eq.~\eqref{eq:Ncross_def} is based on the sign of $\eta_n$, and at smaller $N$, very small values of $\eta_n$ can fluctuate around zero and may lead to spurious sign flips.
	To make the crossing count robust at smaller $N$, one may introduce a finite tolerance $\eta_{\mathrm{tol}}>0$ (as in Eq.~\eqref{eq:eta_tol_introduced}) and discard values with $|\eta_n|\le \eta_{\mathrm{tol}}$ before counting sign flips.
	In the $N=1000$ data presented here, the odd and even Lanczos sub-chains are numerically stable and $\eta_n$ is well resolved over the relevant depth window, so we set $\eta_{\mathrm{tol}}=0$; applying a small positive $\eta_{\mathrm{tol}}$ does not modify the resulting phase boundaries within the resolution of our grid.
	For the same reason we restrict the sign analysis to a recursion window $n\geq n_{\min}=2$ (for an $N$-site chain, the Lanczos recursion generates at most $2N$ coefficients) chosen away from the first couple steps and from the end of the Lanczos run where Lanczos coefficients become unstable (see Appendix~\ref{app:alg_numerical_stability}), where finite-size and roundoff/termination effects are most pronounced. We only count crossings within the numerically stable part of the recursion (see Appendix~\ref{app:alg_numerical_stability}).

	We now use these prescriptions to compute the Krylov subspace based crossing-count phase diagram $N_{\mathrm{cross}}(\alpha,\theta)$ using the boundary seed $\gamma_1$ in the Lanczos recursion, and compare it directly to the edge-bulk gap boundary extracted from the BdG spectrum.
	Figure~\ref{fig:lrk_edge_bulk_phase_diagram} partitions the $(\alpha,\theta)$ plane into a region with no sign changes in the Krylov staggering parameter $\eta_n$ ($N_{\mathrm{cross}}=0$) and a region where nonzero robust sign changes occur ($N_{\mathrm{cross}}\ge 1$).
	The boundary of the $N_{\mathrm{cross}}\ge 1$ region closely follows the edge-bulk gap boundary $\Delta_{\mathrm{edge}}=\Delta_{\mathrm{bulk}}$ obtained from the BdG spectrum, within the resolution of our $(\alpha,\theta)$ grid.
	This correspondence sharpens systematically as the edge-weight threshold $\omega_{\mathrm{edge}}$ is reduced: a conservative choice $\omega_{\mathrm{edge}}=0.5$ already captures the qualitative phase structure, while decreasing $\omega_{\mathrm{edge}}$ to $0.1$ or $0.05$ progressively refines the extracted boundary toward the BdG-derived phase diagram.
	The variation with $\omega_{\mathrm{edge}}$ is most visible in the strong long-range regime ($\alpha < 1$) at large $\theta/\pi$, where algebraically decaying eigenmodes lead to increased finite-size sensitivity in the edge-weight classification.

	The quality of this match depends strongly on the choice of seed operator.
	Boundary seeds, such as $\gamma_1$ used here, maximize the overlap with boundary-localized BdG modes and therefore provide the most sensitive and quantitatively reliable diagnostic of whether the lowest excitation scale is edge dominated or bulk dominated.
	Seeds that remain localized near the boundary but involve a sum of adjacent Majorana operators, e.g.\ $\gamma_1+\gamma_2$, retain clear qualitative sensitivity to the edge-bulk distinction, yet the quantitative agreement with the gap boundary is typically less sharp.
	In contrast, seeds placed deep in the bulk, such as $\gamma_N$ or $\gamma_N+\gamma_{N+1}$ (recalling that an $N$-site chain has $2N$ Majorana modes), couple comparably to left- and right-edge physics and predominantly probe bulk-extended excitations; as a result, the crossing-count signature becomes less tightly locked to the edge-bulk gap boundary and the quantitative agreement degrades further, even though qualitative trends can still be discerned.
	Taken together, these comparisons show that boundary seeds are essential for obtaining a robust and quantitative Krylov subspace diagnostic of boundary-versus-bulk control of the gap, while non-boundary seeds provide at best a weaker, more qualitative probe. Data for the aforementioned additional seeds are provided in Appendix~\ref{app:alt_seeds}.

	\begin{figure}[!t]
	\centering
	\begin{overpic}{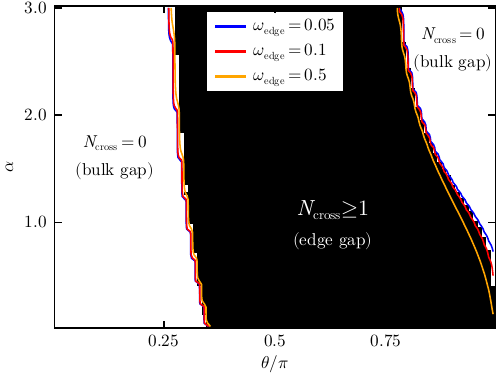}
		\put(12,68){(a) $N=500$}
	\end{overpic}
	\begin{overpic}{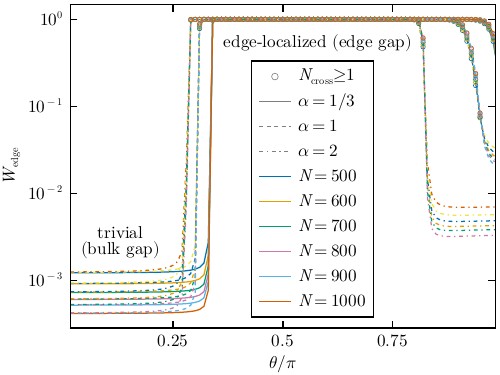}
		\put(16,68){(b)}
	\end{overpic}
	{\phantomsubcaption\label{fig:finite_size_scaling_phase_diag}}
	{\phantomsubcaption\label{fig:finite_size_scaling_edgeloc}}
	\caption{Finite-size robustness of the edge-versus-bulk BdG/Krylov diagnostics at $\epsilon=-0.2$ with boundary seed $\gamma_1$. \subref{fig:finite_size_scaling_phase_diag}~Joint edge-bulk gap and Krylov crossing-count phase diagram at $N=500$ on the same $99\times 99$ $(\theta,\alpha)$ grid used in Fig.~\ref{fig:lrk_edge_bulk_phase_diagram}, with $\ell_{\rm edge}=\lfloor\sqrt{N}\rfloor$; the extracted boundary is visually nearly indistinguishable from the $N=1000$ result in Fig.~\ref{fig:lrk_edge_bulk_phase_diagram}. 
		\subref{fig:finite_size_scaling_edgeloc}~Edge weight $W_{\rm edge}$ versus $\theta/\pi$ for $\alpha=1/3,\,1,\,2$, $\varepsilon = -0.2$ and $N=500,600,700,800,900,1000$ (line style distinguishes $\alpha$ and color distinguishes $N$). In the trivial regime, $W_{\rm edge}$ decreases with increasing $N$, consistent with vanishing boundary weight in the thermodynamic limit, while in the topological regime the curves collapse within visual resolution onto a common finite profile. Open circles mark parameter points with $N_{\rm cross}\geq 1$ [Eq.~\eqref{eq:Ncross_def}]; their onset coincides with the onset of finite $W_{\rm edge}$, showing that the Krylov diagnostic tracks the same boundary-localized physics as the BdG criterion.}
	\label{fig:finite_size_scaling}
\end{figure}

\subsection{Finite-Size Scaling}
\label{app:finite_size_scaling}

To assess the stability of the results presented in Sec.~\ref{subsec:phase diagrams} with respect to system size, we perform two complementary finite-size checks at $\epsilon=-0.2$. First, we repeat the joint BdG edge-bulk gap and Krylov crossing-count phase-diagram analysis at $N=500$, keeping the boundary window $\ell_{\rm edge}=\lfloor\sqrt{N}\rfloor$ and all other operational parameters identical to those used for the $N=1000$ calculations as presented in Fig~\ref{fig:lrk_edge_bulk_phase_diagram}. Second, at fixed $\alpha=1/3,\,1,\,2$, we plot the edge weight $W_{\rm edge}$ as a function of $\theta$ for $N=500,600,700,800,900,1000$ and overlay the parameter points for which $N_{\rm cross}\geq 1$ as defined in Eq.~\eqref{eq:Ncross_def}.

Figure~\ref{fig:finite_size_scaling} summarizes these two checks. \Cref{fig:finite_size_scaling_phase_diag} shows that even the worst-case comparison at $N=500$ yields a phase boundary indistinguishable from the $N=1000$ result used throughout the main text, demonstrating convergence at the system sizes studied in this work. \Cref{fig:finite_size_scaling_edgeloc} shows that in the trivial regime $W_{\rm edge}$ decreases systematically with increasing $N$, consistent with vanishing boundary weight in the thermodynamic limit, whereas in the topological regime the curves for all shown system sizes collapse within visual resolution onto a common finite profile. The onset of finite $W_{\rm edge}$ coincides with the appearance of $N_{\rm cross}\geq 1$, confirming that the Krylov crossing diagnostic tracks the same boundary-localized physics as the BdG-based criterion.

	\section{Conclusion and Outlook}
	\label{sec:conclusion}
	
		This work establishes two central results. First, in the hopping-pairing-balanced short-range Kitaev chain ($\Delta=t$), the single-particle Krylov recursion seeded with the boundary Majorana $\gamma_1$ can be solved exactly: for every recursion depth $n$ and every system size $N$, the Lanczos coefficients satisfy $b_{2n-1}=|\mu|$, $b_{2n}=2|t|$, and $a_n=0$ (Appendix~\ref{app:srk_exact}). The corresponding staggering parameter $\eta_n\equiv\ln(b_{2n-1}/b_{2n})$ is therefore exactly constant, and its sign is an exact topological order parameter: $\eta<0$ if and only if $|\mu|<2|t|$ (topological phase with Majorana zero modes), while $\eta>0$ if and only if $|\mu|>2|t|$ (trivial phase). Thus Kitaev's phase boundary is recovered from a purely Krylov-recursion argument, without invoking winding numbers, band topology, or bulk-gap closure.
		
		Second, in the long-range Kitaev chain at $\epsilon=-0.2$, this exact flat short-range structure is deformed, but its physical content survives. We show that Lanczos coefficients generated from a local boundary seed provide a quantitative diagnostic of whether the lowest excitation gap is controlled by boundary-localized or bulk-extended modes across the full $(\theta,\alpha)$ phase diagram. The key observable is the sign structure of the Krylov staggering parameter $\eta_n$: robust sign changes correlate with edge-dominated low-energy physics, while fixed sign correlates with bulk-dominated regimes. This demonstrates that Lanczos data in quadratic fermionic systems are not merely trivial or featureless, but can encode sharp information about topology, localization, and the spatial origin of the excitation gap. Both results are enabled by an exact single-particle formulation of the operator Lanczos algorithm as derived in this work, which reduces the recursion from exponentially large operator space to a finite-dimensional linear problem and yields machine-precision data for chains of hundreds of sites.

	When the lowest positive BdG mode is classified as edge localized according to the edge-weight criterion in Eq.~\eqref{eq:edge_weight_def}, a boundary seed couples more strongly to that mode, and the Krylov recursion typically shows robust sign changes of $\eta_n$.
	Conversely, when the smallest positive BdG mode is bulk extended, the odd and even subsequences tend to keep a fixed ordering over the stable recursion window and no robust sign crossings occur.
	Importantly, this distinction persists across the short-range to long-range crossover, indicating that the crossing-count diagnostic captures robust edge-versus-bulk physics rather than being tied to fine-tuned model features.

	Our results demonstrate that the diagnostic power of Lanczos data depends critically on seed locality and boundary conditions. Boundary seeds such as the first Majorana operator provide quantitative agreement with gap-based phase boundaries extracted independently from Bogoliubov-de Gennes eigenstates, whereas bulk seeds or seeds spread over multiple sites yield qualitatively weaker signatures. 
	This seed sensitivity underscores a broader principle: operator Krylov diagnostics are most informative when the seed is chosen to couple selectively to the physical degrees of freedom under investigation.
	In the present context, probing boundary-localized edge modes requires boundary-localized seeds.
	This principle is expected to carry over to other settings where Krylov methods are used to diagnose spatial structure and the energy scales that control low-lying excitations.
	
	A natural open question is what sets the \emph{number} of crossings, and whether the count itself carries information beyond the binary distinction of zero versus nonzero crossings.
	Each crossing signals that the odd and even subsequences have swapped order.
	It is possible that different crossings reflect changes in which low-energy excitations couple most strongly to the boundary seed, but clarifying this goes beyond the scope of the present work.
	It would be interesting to test if this crossing count remains meaningful in other models where low-energy physics can switch between edge-localized and bulk-extended excitations, including systems in different Altland-Zirnbauer symmetry classes~\cite{Altland1997Jan,Kitaev2009May}.
	If it does, the count may offer a simple way to further separate edge-dominated regimes by how strongly the relevant edge scale is isolated from the bulk.
	
	The framework developed here opens several immediate extensions. First, the addition of quenched disorder to the long-range Kitaev chain introduces competition between Anderson localization, topological edge physics, and algebraic hybridization induced by power-law pairing. Recent work has shown that disorder can induce reentrant topological behavior and modify the phase diagram of the long-range Kitaev model in nontrivial ways~\cite{Cinnirella2025}. Applying the single-particle Lanczos construction to disordered realizations would reveal whether Krylov staggering remains a reliable diagnostic when disorder competes with boundary localization, and whether disorder-averaged Lanczos coefficients retain edge-bulk sensitivity or exhibit localization-driven signatures analogous to those observed in many-body localized systems~\cite{BallarTrigueros2022}.

	Second, extending the Krylov construction to periodically driven (Floquet) versions of the long-range Kitaev chain would connect our results to recent work on Krylov complexity in time-dependent systems~\cite{Nizami2023}.
	Floquet driving can engineer effective long-range interactions and stabilize dynamical topological phases that have no static counterpart.
	In this direction, it is promising that the operator Krylov space of a broad class of Floquet dynamics can be mapped to an effective one-dimensional Floquet transverse-field Ising model in Krylov space, where edge modes at $0$ and/or $\pi$ quasienergies control long-lived operator dynamics~\cite{Yeh2024Oct}.
	The Arnoldi iteration~\cite{Arnoldi1951} provides a natural generalization of the Lanczos algorithm for Floquet unitaries; a streamlined alternative in Ref.~\cite{Kolganov2025May} exploits orthogonal polynomials on the unit circle to map discrete-time evolution onto a one-dimensional Krylov chain, classifying ergodic and integrable Floquet systems via the asymptotics of Verblunsky coefficients. The Krylov staggering parameter could be adapted to diagnose whether Floquet edge modes remain localized or hybridize with the bulk under driving.
	This extension would also clarify how heating and Floquet prethermalization affect Krylov diagnostics~\cite{Abanin2017Sep,Ikeda2021Oct}, a question of direct relevance to experiments on trapped ions where Floquet protocols are routinely implemented~\cite{Bukov2015Mar,Oka2019Mar,Kiefer2019Nov}.

	Beyond quadratic models, the single-particle closure property exploited here does not extend to interacting systems, where the full many-body operator space must be treated. Nevertheless, the qualitative mechanism underlying Krylov staggering, namely that boundary seeds couple preferentially to edge-localized eigenmodes, is expected to persist in (at least) weakly interacting regimes where edge states remain well defined \cite{koma2022stabilitymajoranaedgezero, Matveeva2024Apr, Hang2025Dec}.
	Numerical studies of interacting Hamiltonians using exact diagonalization or matrix product state methods could test whether the Krylov staggering diagnostic introduced in this work retains its edge-bulk sensitivity when integrability is weakly broken, and whether it remains a viable operational tool in regimes where the eigenstate thermalization hypothesis begins to apply.
	More ambitiously, tensor network techniques could be combined with Krylov recursions to explore operator growth in higher-dimensional systems where long-range interactions induce nontrivial entanglement scaling and modified light-cone structures~\cite{Defenu2023}.
	
	A complementary direction in interacting chains is to extend Krylov-based edge diagnostics to finite temperatures, where edge modes can remain long-lived even when the lifetime is finite.
	Recent work has combined Lanczos-based operator expansions with tensor network representations to access long-time operator dynamics at finite temperature and to extract temperature-dependent decay rates for edge modes in an interacting Kitaev-Hubbard chain~\cite{Tausendpfund2025Jun}.
	Adapting similar ideas to long-range settings could help connect our Krylov staggering-based diagnostics to directly measurable lifetimes and to temperature-dependent crossover scales.

	An experimental protocol to extract Lanczos coefficients from quantum quench statistics has been established for state Krylov complexity~\cite{Pal2023}. Developing analogous protocols for operator dynamics remains an open challenge but would provide direct access to the staggering parameter diagnostic introduced here. The long-range Kitaev chain can be realized in trapped-ion quantum simulators and Rydberg atom arrays, providing natural testbeds for such measurements. In trapped-ion platforms, power-law spin-spin interactions with tunable exponent have been demonstrated~\cite{Richerme2014,Jurcevic2014,Foss_Feig2025Mar}, and the mapping between fermionic and spin models via Jordan-Wigner transformation places the long-range Kitaev chain within experimental reach. Rydberg atom arrays with tunable long-range interactions provide another promising platform~\cite{Browaeys2020,Cheng2024}. Similarly, cavity-mediated interactions in ultracold atomic gases can engineer flat or algebraically decaying couplings~\cite{Ritsch2013,Schlawin2019Sep}, enabling direct simulation of the pairing terms studied here. In the short-range limit, minimal Kitaev chains consisting of two coupled quantum dots have been realized in semiconductor-superconductor hybrid nanowires~\cite{Wang2022,Dvir2023}, with recent work demonstrating enhanced Majorana stability in three-site chains~\cite{Bordin2024}. While these implementations access only nearest-neighbor couplings, they provide a complementary bottom-up route to engineering controllable topological systems and probing Majorana physics at the few-site level. Together, these platforms position Krylov diagnostics as practical probes of edge-bulk competition and localization physics once operator-growth measurement protocols become available in analog quantum simulators.

	\section*{Data and code availability}
	
	All data and code used for data generation are available on Zenodo~\cite{this_zenodo}.
	
	\section*{Acknowledgments}
	This work is funded by the Deutsche Forschungsgemeinschaft (DFG, German Research Foundation) – 436382789, 493420525, 499180199; via FOR 5522 and large-equipment grants (GOEGrid cluster).
	
	\bibliography{refs.bib}
	
	
	\appendix
	\clearpage
	
	\section{Majorana commutator closure}
	\label{app:majorana_commutator}
	
	This appendix provides a self contained derivation of the commutator identity
	\begin{equation}
		[H_{\mathrm{LRK}},\gamma_\ell]= i\sum_{m=1}^{2N}\mathcal{H}_{M,m\ell}\gamma_m,
		\label{eq:comm_H_gamma_app}
	\end{equation}
	used in Sec.~\ref{subsec:model_liouvillian} to obtain a closed single-particle equation of motion for operators linear in Majoranas.
	
	\subsection{Trace identities}
	\label{app:majorana_traces}
	
	The Majorana operators satisfy $\gamma_\mu^\dagger=\gamma_\mu$ and $\{\gamma_\mu,\gamma_\nu\}=\delta_{\mu\nu}$.
	With the normalization in Eq.~\eqref{eq:majorana_def}, one has $\gamma_\mu^2=\frac{1}{2}$.
	Hence, on a $2^N$ dimensional fermionic Hilbert space,
	\begin{equation}
		\mathrm{Tr}(\gamma_\mu)=0,
		\qquad
		\mathrm{Tr}(\gamma_\mu\gamma_\nu)=2^{N-1}\delta_{\mu\nu}.
	\end{equation}
	The second identity follows from $\mathrm{Tr}(\gamma_\mu^2)=\mathrm{Tr}(\mathds{1}/2)=2^{N-1}$ and from $\mathrm{Tr}(\gamma_\mu\gamma_\nu)= -\mathrm{Tr}(\gamma_\nu\gamma_\mu)$ for $\mu\neq\nu$.
	
	\subsection{Elementary commutator identity}
	\label{app:commutator_identity}
	
	We use the operator identity
	\begin{equation}
		[\gamma_j\gamma_k,\gamma_\ell]
		=\gamma_j\{\gamma_k,\gamma_\ell\}-\{\gamma_j,\gamma_\ell\}\gamma_k.
		\label{eq:gamma_comm_identity}
	\end{equation}
	To verify it, expand and reorder using anticommutators:
	\begin{align}
		[\gamma_j\gamma_k,\gamma_\ell]
		&=\gamma_j\gamma_k\gamma_\ell-\gamma_\ell\gamma_j\gamma_k \nonumber\\
		&=\gamma_j(\gamma_k\gamma_\ell)-(\gamma_\ell\gamma_j)\gamma_k \nonumber\\
		&=\gamma_j(\{\gamma_k,\gamma_\ell\}-\gamma_\ell\gamma_k)-(\{\gamma_\ell,\gamma_j\}-\gamma_j\gamma_\ell)\gamma_k \nonumber\\
		&=\gamma_j\{\gamma_k,\gamma_\ell\}-\{\gamma_\ell,\gamma_j\}\gamma_k,
	\end{align}
	which is Eq.~\eqref{eq:gamma_comm_identity}.
	With $\{\gamma_\mu,\gamma_\nu\}=\delta_{\mu\nu}$, this immediately yields
	\begin{equation}
		[\gamma_i\gamma_j,\gamma_\ell]=\gamma_i\,\delta_{j\ell}-\delta_{i\ell}\,\gamma_j.
		\label{eq:gamma_gamma_comm}
	\end{equation}
	
	\subsection{\texorpdfstring{Derivation of $[H_{\mathrm{LRK}},\gamma_\ell]$}{Derivation of [H LRK, gamma l]}}
	\label{app:derivation_commutator}
	
	Starting from the quadratic Majorana form
	\begin{equation}
		H_{\mathrm{LRK}}=\frac{i}{2}\sum_{i,j=1}^{2N}\mathcal{H}_{M,ij}\gamma_i\gamma_j,
		\label{eq:H_majorana_app}
	\end{equation}
	where $\mathcal{H}_M$ is real and antisymmetric, $\mathcal{H}_{M,ji}=-\mathcal{H}_{M,ij}$, we compute
	\begin{align}
		[H_{\mathrm{LRK}},\gamma_\ell]
		&=\frac{i}{2}\sum_{i,j=1}^{2N}\mathcal{H}_{M,ij}[\gamma_i\gamma_j,\gamma_\ell] \nonumber\\
		&=\frac{i}{2}\sum_{i,j=1}^{2N}\mathcal{H}_{M,ij}\left(\gamma_i\delta_{j\ell}-\delta_{i\ell}\gamma_j\right) \nonumber\\
		&=\frac{i}{2}\sum_{i=1}^{2N}\mathcal{H}_{M,i\ell}\gamma_i-\frac{i}{2}\sum_{j=1}^{2N}\mathcal{H}_{M,\ell j}\gamma_j.
	\end{align}
	Relabel $j\to i$ in the second term and use antisymmetry $\mathcal{H}_{M,\ell i}=-\mathcal{H}_{M,i\ell}$ to obtain
	\begin{equation}
		[H_{\mathrm{LRK}},\gamma_\ell]
		=\frac{i}{2}\sum_{i=1}^{2N}\mathcal{H}_{M,i\ell}\gamma_i+\frac{i}{2}\sum_{i=1}^{2N}\mathcal{H}_{M,i\ell}\gamma_i
		=i\sum_{i=1}^{2N}\mathcal{H}_{M,i\ell}\gamma_i,
	\end{equation}
	which is Eq.~\eqref{eq:comm_H_gamma_app}.

	\subsection{Closure for linear operators}
	\label{app:closure_linear_ops}
	
	Consider any generic operator $\mathcal{O}=\sum_{k=1}^{2N}u_k\gamma_k$.
	Using Eq.~\eqref{eq:comm_H_gamma_app},
	\begin{align}
		[H_{\mathrm{LRK}},\mathcal{O}]
		&=\sum_{k=1}^{2N}u_k[H_{\mathrm{LRK}},\gamma_k] \nonumber\\
		&= i\sum_{k=1}^{2N}u_k\sum_{m=1}^{2N}\mathcal{H}_{M,mk}\gamma_m
		= i\sum_{m=1}^{2N}(\mathcal{H}_M u)_m\,\gamma_m,
		\label{eq:commutator_linear_op_app}
	\end{align}
	showing that the commutator maps the linear Majorana subspace to itself.

	\section{Algorithmic details}
	\label{app:algorithm_details}
	
	\subsection{Inner product on the linear Majorana subspace}
	\label{app:alg_inner_product}
	
	Consider operators $\mathcal{O}_v=\sum_{\mu}v_\mu\gamma_\mu$ and $\mathcal{O}_w=\sum_{\mu}w_\mu\gamma_\mu$.
	Using $\mathrm{Tr}(\gamma_\mu\gamma_\nu)=2^{N-1}\delta_{\mu\nu}$ for the Majorana normalization adopted in Eq.~\eqref{eq:majorana_def}, the infinite-temperature Hilbert-Schmidt inner product becomes
	\begin{equation}
		\langle \mathcal{O}_v,\mathcal{O}_w\rangle_{\mathrm{HS}}
		=\frac{1}{2^N}\mathrm{Tr}\!\left(\mathcal{O}_v^\dagger \mathcal{O}_w\right)
		=\frac{1}{2}\sum_{\mu=1}^{2N} v_\mu^\ast w_\mu
		=\frac{1}{2}\,v^\dagger w.
	\end{equation}
	Thus, within the linear Majorana sector, the Euclidean inner product differs from the Hilbert-Schmidt inner product only by an overall constant factor.
	
	\subsection{\texorpdfstring{Proof that $a_n=0$ for $\mathcal{O}(0)^\dagger=\pm\mathcal{O}(0)$}{Proof that a_n = 0 for O\^dagger(0) = +/- O(0)}}
	\label{app:alg_an_zero}
	
	Let $H$ be Hermitian and define the Hilbert-Schmidt inner product
	$\langle \mathcal{A},\mathcal{B}\rangle_{\mathrm{HS}}=\mathrm{Tr}(\mathcal{A}^\dagger\mathcal{B})/2^N$.
	For any operator $\mathcal{A}$ satisfying $\mathcal{A}^\dagger=\pm \mathcal{A}$, we have
	\begin{equation}
		\langle \mathcal{A},[H,\mathcal{A}]\rangle_{\mathrm{HS}}=0.
		\label{eq:HS_A_comm_A_zero}
	\end{equation}
	Using cyclicity of the trace,
	\begin{align}
		\mathrm{Tr}\!\left(\mathcal{A}^\dagger[H,\mathcal{A}]\right)
		&=\mathrm{Tr}\!\left(\mathcal{A}^\dagger H\mathcal{A}-\mathcal{A}^\dagger\mathcal{A}H\right) \nonumber\\
		&=\mathrm{Tr}\!\left(H\mathcal{A}\mathcal{A}^\dagger-\mathcal{A}^\dagger\mathcal{A}H\right).
	\end{align}
	If $\mathcal{A}^\dagger=\mathcal{A}$ (Hermitian), then $\mathcal{A}\mathcal{A}^\dagger=\mathcal{A}^2=\mathcal{A}^\dagger\mathcal{A}$ and the terms cancel.
	If $\mathcal{A}^\dagger=-\mathcal{A}$ (anti-Hermitian), then $\mathcal{A}\mathcal{A}^\dagger=-\mathcal{A}^2=\mathcal{A}^\dagger\mathcal{A}$ and the cancellation is identical.
	This proves Eq.~\eqref{eq:HS_A_comm_A_zero}.
	
	Running the operator Lanczos recursion for $\mathcal{L}=[H,\cdot]$ with a Hermitian or anti-Hermitian seed implies that all normalized Lanczos operators satisfy $\mathcal{O}_n^\dagger=\pm\mathcal{O}_n$.
	Therefore,
	\begin{equation}
		a_n=\langle \mathcal{O}_n,\mathcal{L}\mathcal{O}_n\rangle_{\mathrm{HS}}=0
		\qquad\text{for all }n.
	\end{equation}
	Via Appendix~\ref{app:alg_inner_product}, the same conclusion applies to the coefficient-space implementation used in the main text.

	\subsection{Derivation of the Krylov subspace evolution equation}
	\label{app:alg_krylov_ode}
	
	The tridiagonal Lanczos matrix $T$ is defined in Eq.~\eqref{eq:T_def_local} and has entries $T_{nn}=a_n$ and $T_{n,n+1}=T_{n+1,n}=b_{n+1}$.
	For the Hermitian seed (as is considered in this work), $a_n=0$ and $T$ is a real symmetric matrix with a vanishing diagonal.
	
	The coefficient vector evolves according to (using Eq.~\eqref{eq:single_particle_liouvillian} in Eq.~\eqref{eq:eom_coeffs_main})
	\begin{equation}
		\frac{d u(t)}{dt} = i L_{\mathrm{sp}}\, u(t),
		\qquad
		u(0)=v_0.
		\label{eq:a_time_evolution}
	\end{equation}
	Let $V=(v_0,\ldots,v_{\mathcal{K}-1})$ denote the matrix whose columns are the Lanczos vectors.
	We expand $u(t)$ in the Krylov basis as
	\begin{equation}
		u(t)=\sum_{n=0}^{\mathcal{K}-1}\phi_n(t)\,v_n = V\phi(t),
		\quad
		\phi(t)\equiv(\phi_0(t),\ldots,\phi_{\mathcal{K}-1}(t))^T.
		\label{eq:krylov_expansion}
	\end{equation}
	Since $v_0=u(0)$ is normalized, the initial Krylov amplitudes are
	\begin{equation}
		\phi(0)=e_0\equiv(1,0,\ldots,0)^T.
		\label{eq:phi0}
	\end{equation}
	Differentiating Eq.~\eqref{eq:krylov_expansion} gives $\dot{u}(t)=V\dot{\phi}(t)$.
	Using $\dot{u}(t)=iL_{\mathrm{sp}}u(t)$ from Eq.~\eqref{eq:a_time_evolution} and left-multiplying by $V^\dagger$, we obtain
	\begin{equation}
		\dot{\phi}(t)= i \left(V^\dagger L_{\mathrm{sp}} V\right)\phi(t) = iT\phi(t),
	\end{equation}
	or equivalently,
	\begin{equation}
		\frac{d\phi(t)}{dt} = i T\,\phi(t),
		\qquad
		\phi(0)=e_0,
		\label{eq:krylov_schrodinger_main}
	\end{equation}
	which is unitary in the finite-dimensional Krylov space.

	\subsection{Reorthogonalization and stability checks}
	\label{app:alg_numerical_stability}
	
	In finite precision arithmetic, the Lanczos basis is bound to loose orthogonality for large number of Lanczos steps $n$, leading to strong numerical instability.
	This unfortunate fact is almost impossible to get around; even when using sophisticated re-orthogonalization routines as they are used across the literature.
	In fact, often the introduction of a re-orthogonalization routine brings in its own numerical instability problems.
	Further, even when we are able to show that the Lanczos basis is perfectly orthogonal (up to machine precision), this does not imply that the sequence of Lanczos coefficients is accurate.
	For example, it was shown in Ref.~\cite{Eckseler2025} that in finite precision arithmetic, Lanczos sequences can escape disconnected subspaces, leading to incorrect sequences at large $n$.
	This shows that one always has to be careful when calculating Lanczos coefficients and that it is almost impossible to show that any calculated sequence of Lanczos coefficients is exact for large $n$.
	
	For this manuscript, the single-particle Lanczos algorithm is particularly well-conditioned for the model we have considered in formally equivalent, yet numerically distinct formulations: Nambu and Majorana formalisms.
	Here we explain the all the checks we impose to obtain Lanczos coefficients which are convincingly accurate and rigorous. 
	
	To achieve this, we use partial reorthogonalization after forming $w_n$ in the Lanczos routine:
	\begin{equation}
		w_n \leftarrow w_n - \sum_{j=0}^{n}\langle v_j,w_n\rangle\,v_j\Theta(|\langle v_j,w_n\rangle| - p).
	\end{equation}
	In words, we reorthogonalize only whenever the overlap $|\langle v_j, w_n\rangle| > p$ (here $\Theta(\cdot)$ is the Heaviside function), where we choose $p=10^{-10}$ across all sequences of Lanczos coefficients analyzed in this manuscript.
	This method strikes a balance between the inherent instability of the Lanczos algorithm and the instability introduced by orthogonalizing a basis in a Gram-Schmidt-like fashion, as is well established in the field.
	In the literature, it is often recommended to perform this step twice, but we do not find any improvement for our simulation.
	
	While the partial reorthogonalization technique is best-practice in the field, the accuracy of the Lanczos sequence is still not guaranteed.
	Therefore, we have performed multiple checks regarding the accuracy of the presented sequences of Lanczos sequences.
	Firstly, we monitor the loss of orthogonality in the Lanczos basis vectors $\{ v_i\}$ using
	\begin{equation}
		\varepsilon_n \equiv \max_{0\le i<n}|\langle v_i,v_n\rangle|.
	\end{equation}
	If we find that $\max_n \varepsilon_n > 10^{-7}$ we regard the rest of the sequence as unstable and discard subsequent Lanczos coefficients.
	It is important to note here, that we also terminate the Lanczos sequence whenever $b_n \le 10^{-7}$.
	
	As discussed previously, the orthogonality of the basis does not guarantee the correctness of the sequence of Lanczos coefficients.
	Therefore, we perform the Lanczos algorithm with partial reorthogonalization using two formally equivalent but numerically distinct algorithms, namely the Lanczos sequence in the Majorana representation as well as in the Nambu representation.
	Formally these algorithms are exactly equivalent, however numerically they are quite different.
	Foremost, the Lanczos algorithm in the Nambu representation uses entirely real arithmetic, while the Majorana algorithm needs imaginary values.
	This ensures that we have a rigorous benchmark, as we don't expect the two algorithms to coincide in their results whenever the Lanczos algorithm is highly unstable.
	In practice, we impose that the difference between the sequences of the two algorithms must not be larger than $10^{-7}$.

\section{Exact Thermodynamic Limit Lanczos Coefficients for the Short-Range Kitaev Chain}
\label{app:srk_exact}

The long-range Kitaev Hamiltonian~\eqref{eq:hamiltonian_long_range_kitaev} reduces, in the limit $\alpha\to\infty$, to a model with nearest-neighbor couplings only. Retaining only the $|i-j|=1$ terms yields
\begin{equation}
	H_{\mathrm{SRK}} = \sum_{j=1}^{N-1}\!\bigl[{t}(c_j^\dagger c_{j+1}+\mathrm{h.c.}) + \Delta(c_j^\dagger c_{j+1}^\dagger +\mathrm{h.c.})\bigr] - \mu\sum_{j=1}^{N} c_j^\dagger c_j,
	\label{eq:H_srk_app}
\end{equation}
with $t = \sin\theta$, $\Delta = (1+\epsilon)\sin\theta$, and $\mu = -2\cos\theta$ as defined in Sec.~\ref{subsec:model_hamiltonian}. This is the standard short-range Kitaev chain as originally proposed in Ref.~\cite{Kitaev2001}. Throughout this appendix we set $\epsilon = 0$ for analytical tractability, i.e.
\begin{equation}
	\Delta = t,
	\label{eq:sweet_spot_app}
\end{equation}
so that $\mu$ and $t$ are the only two independent parameters, $t + \Delta = 2t$, and $t - \Delta = 0$.

The central result is that the off-diagonal Lanczos coefficients produced by the single-particle Krylov algorithm (Sec.~\ref{sec:algorithm}), seeded with $\gamma_1$ (Eq.~\eqref{eq:seed_operator_gamma1}), are perfectly flat in alternating pairs for every $n$ and every $\mu$:
\begin{equation}
	\boxed{b_{2n-1} = |\mu|, \quad b_{2n} = 2|t|, \quad a_n = 0 \qquad \forall n.}
	\label{eq:exact_bn_app}
\end{equation}
The Krylov staggering parameter (Eq.~\eqref{eq:etadef_main}) is therefore constant:
\begin{equation}
	\eta_n = \ln\!\left(\frac{b_{2n-1}}{b_{2n}}\right) = \ln\!\left(\frac{|\mu|}{2|t|}\right), \qquad \forall\, n.
	\label{eq:eta_exact_app}
\end{equation}
Numerical confirmation is shown in Fig.~\ref{fig:app_exact_srk}.

\subsection{Majorana representation and the matrix \texorpdfstring{$\mathcal{H}_M$}{HM}}
\label{app:srk_HM}

\paragraph{Majorana basis and coefficient vectors.}
Following Eqs.~\eqref{eq:majorana_def}--\eqref{eq:inverse_majorana}, each site $j$ contributes two Majorana operators,
\begin{equation}
	\gamma_{2j-1} = \frac{c_j + c_j^\dagger}{\sqrt{2}}, \qquad \gamma_{2j} = \frac{c_j^\dagger - c_j}{i\sqrt{2}},
	\label{eq:gamma_split_app}
\end{equation}
satisfying $\gamma_\mu^\dagger = \gamma_\mu$ and $\{\gamma_\mu, \gamma_\nu\} = \delta_{\mu\nu}$. We call $\gamma_{2j-1}$ the \emph{A-type} and $\gamma_{2j}$ the \emph{B-type} Majorana on site $j$. The Hamiltonian is written as $H_\mathrm{SRK} = \frac{i}{2}\sum_{\mu\nu} \mathcal{H}_{M,\mu\nu}\,\gamma_\mu\gamma_\nu$ (Eq.~\eqref{eq:H_majorana}), with $\mathcal{H}_M$ real and antisymmetric: $\mathcal{H}_{M,\mu\nu} = -\mathcal{H}_{M,\nu\mu}$. Coefficient vectors $v \in \mathbb{C}^{2N}$ are ordered as $v = (v_{\gamma_1}, v_{\gamma_2}, \ldots, v_{\gamma_{2N}})^T$, and $\mathbf{e}_\mu$ denotes the unit vector with a $1$ at position $\mu$ (1-indexed). The Krylov recursion is seeded by $v_0 = \mathbf{e}_{\gamma_1}$ (Eq.~\eqref{eq:seed_operator_gamma1}).

\paragraph{Computing the entries of \texorpdfstring{$\mathcal{H}_M$}{HM}.}
Because $\mathcal{H}_M$ is antisymmetric and the double sum $\frac{i}{2}\sum_{\mu\nu}\mathcal{H}_{M,\mu\nu}\gamma_\mu\gamma_\nu$ counts each ordered pair $(\mu,\nu)$ and $(\nu,\mu)$ separately, we expand it for a fixed pair $\mu < \nu$:
\begin{equation}
	\begin{aligned}
		\frac{i}{2}\bigl(\mathcal{H}_{M,\mu\nu}\,\gamma_\mu\gamma_\nu + \mathcal{H}_{M,\nu\mu}\,\gamma_\nu\gamma_\mu\bigr) =& \frac{i}{2}\,\mathcal{H}_{M,\mu\nu}\bigl(\gamma_\mu\gamma_\nu - \gamma_\nu\gamma_\mu\bigr) \\
		=& i\,\mathcal{H}_{M,\mu\nu}\,\gamma_\mu\gamma_\nu,
	\end{aligned}
	\label{eq:dictionary_derive_app}
\end{equation}
where we used antisymmetry $\mathcal{H}_{M,\nu\mu}=-\mathcal{H}_{M,\mu\nu}$ and the anticommutation $\gamma_\nu\gamma_\mu = -\gamma_\mu\gamma_\nu$ (valid for $\mu\neq\nu$ since $\{\gamma_\mu,\gamma_\nu\}=0$). Therefore a term $iA\,\gamma_\mu\gamma_\nu$ appearing in $H$ with $\mu < \nu$ implies
\begin{equation}
	\mathcal{H}_{M,\mu\nu} = A, \qquad \mathcal{H}_{M,\nu\mu} = -A.
	\label{eq:HM_dictionary_app}
\end{equation}
We use this dictionary throughout. Inverting Eq.~\eqref{eq:gamma_split_app}: $c_j = (\gamma_{2j-1} - i\gamma_{2j})/\sqrt{2}$ and $c_j^\dagger = (\gamma_{2j-1} + i\gamma_{2j})/\sqrt{2}$.

\textit{On-site term.} Expanding directly:
\begin{align}
	c_j^\dagger c_j &= \frac{(\gamma_{2j-1}+i\gamma_{2j})(\gamma_{2j-1}-i\gamma_{2j})}{2} \notag\\ &= \frac{\gamma_{2j-1}^2 - i\gamma_{2j-1}\gamma_{2j} + i\gamma_{2j}\gamma_{2j-1} + \gamma_{2j}^2}{2},
\end{align}
where $\gamma_\mu^2 = \frac{1}{2}$ (from $\{\gamma_\mu,\gamma_\mu\}=1$) and $\gamma_{2j}\gamma_{2j-1} = -\gamma_{2j-1}\gamma_{2j}$ (from $\{\gamma_{2j-1},\gamma_{2j}\}=0$). Substituting and dropping the constant $\frac{1}{2}$ (which commutes with everything and does not affect the Krylov recursion):
\begin{equation}
	c_j^\dagger c_j = \frac{1}{2} - i\gamma_{2j-1}\gamma_{2j} \quad\Longrightarrow\quad -\mu c_j^\dagger c_j \;\longrightarrow\; +i\mu\,\gamma_{2j-1}\gamma_{2j}.
	\label{eq:onsite_majorana_app}
\end{equation}
By the dictionary~\eqref{eq:HM_dictionary_app} with $A = \mu$ and $(\mu,\nu) = (2j-1,2j)$:
\begin{equation}
	\mathcal{H}_{M,\,2j-1,\,2j} = +\mu, \qquad \mathcal{H}_{M,\,2j,\,2j-1} = -\mu.
	\label{eq:HM_onsite_app}
\end{equation}

\textit{Hopping term.} Setting $A\equiv\gamma_{2j-1}$, $B\equiv\gamma_{2j}$, $C\equiv\gamma_{2j+1}$, $D\equiv\gamma_{2j+2}$ for brevity, expand:
\begin{equation}
	c_j^\dagger c_{j+1} + c_{j+1}^\dagger c_j = \frac{(A+iB)(C-iD)+(C+iD)(A-iB)}{2}.
\end{equation}
The same-sublattice products $AC+CA$ and $BD+DB$ vanish by anticommutation. The cross terms give $BC-CB = 2BC$ and $AD-DA = 2AD$, so
\begin{equation}
	c_j^\dagger c_{j+1} + c_{j+1}^\dagger c_j = i\!\left[\gamma_{2j}\gamma_{2j+1} - \gamma_{2j-1}\gamma_{2j+2}\right].
	\label{eq:hop_expand_app}
\end{equation}
Multiplying by $+t$ (the sign in $H_\mathrm{SRK}$):
\begin{equation}
	+t(c_j^\dagger c_{j+1}+\mathrm{h.c.}) = +it\!\left[\gamma_{2j}\gamma_{2j+1} - \gamma_{2j-1}\gamma_{2j+2}\right].
	\label{eq:hopping_majorana_app}
\end{equation}

\textit{Pairing term.} The analogous expansion of $c_j^\dagger c_{j+1}^\dagger + c_{j+1}c_j$ yields:
\begin{equation}
	c_j^\dagger c_{j+1}^\dagger + c_{j+1}c_j = +i\!\left[\gamma_{2j-1}\gamma_{2j+2} + \gamma_{2j}\gamma_{2j+1}\right],
	\label{eq:pair_expand_app}
\end{equation}
which differs from the hopping expansion~\eqref{eq:hop_expand_app} in the sign of the $\gamma_{2j-1}\gamma_{2j+2}$ term. Note also the key identity $c_j^\dagger c_{j+1}^\dagger + \mathrm{h.c.} = -(c_jc_{j+1}+\mathrm{h.c.})$, so writing the pairing as $c^\dagger c^\dagger$ rather than $cc$ is not a matter of notation but fixes the sign. Multiplying by $+\Delta = +t$:
\begin{equation}
	+t(c_j^\dagger c_{j+1}^\dagger+\mathrm{h.c.}) = +it\!\left[\gamma_{2j-1}\gamma_{2j+2} + \gamma_{2j}\gamma_{2j+1}\right].
	\label{eq:pairing_majorana_app}
\end{equation}

\textit{Adding hopping and pairing.} The $\gamma_{2j-1}\gamma_{2j+2}$ terms cancel between Eqs.~\eqref{eq:hopping_majorana_app} and \eqref{eq:pairing_majorana_app}:
\begin{equation}
	\begin{aligned}
		+t&(c_j^\dagger c_{j+1}+\mathrm{h.c.}) + t(c_j^\dagger c_{j+1}^\dagger+\mathrm{h.c.}) \\
		&= +it\bigl[-\gamma_{2j-1}\gamma_{2j+2}+\gamma_{2j-1}\gamma_{2j+2}\bigr] + 2it\,\gamma_{2j}\gamma_{2j+1} \\
		&= +2it\,\gamma_{2j}\gamma_{2j+1}.
	\end{aligned}
	\label{eq:hop_pair_combined_app}
\end{equation}
This cancellation is precisely the content of $\Delta = t$: at general $\Delta$, the $\gamma_{2j-1}\gamma_{2j+2}$ coefficient is $i(\Delta-t)$, which vanishes only when $\Delta=t$. By the dictionary~\eqref{eq:HM_dictionary_app} with $A = +2t$ and $(\mu,\nu)=(2j, 2j+1)$:
\begin{equation}
	\mathcal{H}_{M,\,2j,\,2j+1} = +2t, \qquad \mathcal{H}_{M,\,2j+1,\,2j} = -2t.
	\label{eq:HM_intersite_app}
\end{equation}

\textit{Summary.} The nonzero independent entries of $\mathcal{H}_M$ at $\Delta=t$ are (plus antisymmetric partners):
\begin{center}
	\renewcommand{\arraystretch}{1.4}
	\begin{tabular}{lll}
		\hline
		Entry & Value & Origin \\
		\hline
		$\mathcal{H}_{M,\,2j-1,\,2j}$ & $+\mu$ & on-site \\[2pt]
		$\mathcal{H}_{M,\,2j,\,2j+1}$ & $+(t{+}\Delta) \to {+}2t$ & hopping $+$ pairing ($\Delta{=}t$) \\[2pt]
		$\mathcal{H}_{M,\,2j-1,\,2j+2}$ & $(\Delta{-}t) \to 0$ & cancel at $\Delta{=}t$ \\
		\hline
	\end{tabular}
\end{center}
The third entry vanishes exactly because $\Delta = t$: at general $\Delta$, the inter-site result from Eqs.~\eqref{eq:hopping_majorana_app}--\eqref{eq:pairing_majorana_app} is
\begin{equation}
	\mathcal{H}_{M,\,2j-1,\,2j+2} = +(\Delta-t), \qquad \mathcal{H}_{M,\,2j,\,2j+1} = +(t+\Delta),
	\label{eq:HM_general_Delta_app}
\end{equation}
so both entries recover their sweet-spot values $0$ and $+2t$ when $\Delta=t$. The matrix $\mathcal{H}_M$ is therefore \emph{tridiagonal}:
\begin{equation}
	\boxed{\mathcal{H}_{M,\,2j-1,\,2j} = +\mu, \qquad \mathcal{H}_{M,\,2j,\,2j+1} = +2t.}
	\label{eq:HM_tridiag_app}
\end{equation}
Pictorially, $\mathcal{H}_M$ is the adjacency matrix of a 1D chain with alternating bond strengths:
\begin{equation}
	\gamma_1 \;\xrightarrow{|\mu|}\; \gamma_2 \;\xrightarrow{2|t|}\; \gamma_3 \;\xrightarrow{|\mu|}\; \gamma_4 \;\xrightarrow{2|t|}\; \gamma_5 \;\xrightarrow{|\mu|}\; \cdots
	\label{eq:chain_picture_app}
\end{equation}

\paragraph{Action of \texorpdfstring{$L_\mathrm{sp}$}{Lsp} on basis vectors.}
The single-particle Liouvillian $L_\mathrm{sp} = i\mathcal{H}_M$ is Hermitian. For the matrix-vector product $L_\mathrm{sp}\,\mathbf{e}_\mu$, we need \emph{column} $\mu$ of $\mathcal{H}_M$, i.e.\ the entries $\mathcal{H}_{M,\nu\mu}$ for all $\nu$; these are the antisymmetric partners of the primary entries in Eq.~\eqref{eq:HM_tridiag_app}:
\begin{subequations}
	\label{eq:Lsp_actions_app}
	\begin{align}
		L_\mathrm{sp}\,\mathbf{e}_{\gamma_1} &= -i\mu\,\mathbf{e}_{\gamma_2}, \label{eq:Lsp_gamma1_app}\\
		L_\mathrm{sp}\,\mathbf{e}_{\gamma_{2j}} &= +i\mu\,\mathbf{e}_{\gamma_{2j-1}} - 2it\,\mathbf{e}_{\gamma_{2j+1}}, \qquad j \leq N-1, \label{eq:Lsp_B_app}\\
		L_\mathrm{sp}\,\mathbf{e}_{\gamma_{2j+1}} &= +2it\,\mathbf{e}_{\gamma_{2j}} - i\mu\,\mathbf{e}_{\gamma_{2j+2}}, \qquad j \geq 1. \label{eq:Lsp_A_app}
	\end{align}
\end{subequations}
Each formula follows by multiplying $i$ with the relevant column of $\mathcal{H}_M$. For column $1$ (boundary, $\gamma_1$): the only nonzero entry is $\mathcal{H}_{M,2,1} = -\mu$ (antisymmetric partner of $\mathcal{H}_{M,1,2}=+\mu$), giving $L_\mathrm{sp}\,\mathbf{e}_{\gamma_1} = i(-\mu)\mathbf{e}_{\gamma_2} = -i\mu\,\mathbf{e}_{\gamma_2}$. For column $2j$ (B-type on site $j$, $j\leq N-1$): the two nonzero entries are $\mathcal{H}_{M,2j-1,\,2j} = +\mu$ (primary on-site rule) and $\mathcal{H}_{M,2j+1,\,2j} = -2t$ (antisymmetric partner of the inter-site primary $\mathcal{H}_{M,2j,\,2j+1}=+2t$); multiplying by $i$: $i(+\mu)\mathbf{e}_{\gamma_{2j-1}} + i(-2t)\mathbf{e}_{\gamma_{2j+1}} = +i\mu\,\mathbf{e}_{\gamma_{2j-1}} - 2it\,\mathbf{e}_{\gamma_{2j+1}}$, giving Eq.~\eqref{eq:Lsp_B_app}. For column $2j+1$ (A-type on site $j+1$, $j\geq 1$): the two nonzero entries are $\mathcal{H}_{M,2j,\,2j+1} = +2t$ (primary inter-site rule) and $\mathcal{H}_{M,2j+2,\,2j+1} = -\mu$ (antisymmetric partner of the on-site primary $\mathcal{H}_{M,2j+1,\,2j+2}=+\mu$); multiplying by $i$: $i(+2t)\mathbf{e}_{\gamma_{2j}} + i(-\mu)\mathbf{e}_{\gamma_{2j+2}} = +2it\,\mathbf{e}_{\gamma_{2j}} - i\mu\,\mathbf{e}_{\gamma_{2j+2}}$, giving Eq.~\eqref{eq:Lsp_A_app}.

\subsection{Proof by induction}
\label{app:srk_proof}

\noindent\textbf{Claim.} For $H_\mathrm{SRK}$ at $\Delta = t$, with seed $v_0 = \mathbf{e}_{\gamma_1}$, and without loss of generality $t, \mu > 0$, the Lanczos recursion
\begin{equation}
	\begin{aligned} w_n =& L_\mathrm{sp}\,v_{n-1} - a_{n-1}v_{n-1} - b_{n-1}v_{n-2}, \\ b_n =& \|w_n\|, \qquad v_n = \frac{w_n}{b_n}, \end{aligned}
	\label{eq:recurrence_app}
\end{equation}
produces, for all $n = 0, 1, 2, \ldots, 2N-1$:
\begin{equation}
	v_n = (-i)^n\,\mathbf{e}_{\gamma_{n+1}},
	\label{eq:vn_claim_app}
\end{equation}
with $a_n = 0$ for all $n$, $b_{2n-1} = |\mu|$, and $b_{2n} = 2|t|$. Note that $(-i)^n$ cycles through $+1, -i, -1, +i$ with period 4, so even-indexed vectors alternate in real sign ($v_{4k}=+\mathbf{e}_{\gamma_{4k+1}}$, $v_{4k+2}=-\mathbf{e}_{\gamma_{4k+3}}$) while odd-indexed vectors alternate in imaginary sign ($v_{4k+1}=-i\,\mathbf{e}_{\gamma_{4k+2}}$, $v_{4k+3}=+i\,\mathbf{e}_{\gamma_{4k+4}}$). The Lanczos coefficients $b_n$ and $a_n$ depend only on norms and inner products and are therefore insensitive to these phase factors.

\subsubsection*{\texorpdfstring{Base case ($v_0 \to v_1$)}{Base case}}

Starting from $v_0 = \mathbf{e}_{\gamma_1}$ (real unit vector, consistent with $(-i)^0=1$):
\begin{equation}
	L_\mathrm{sp}\,v_0 \stackrel{\eqref{eq:Lsp_gamma1_app}}{=} -i\mu\,\mathbf{e}_{\gamma_2}.
\end{equation}
The diagonal coefficient vanishes because $\mathbf{e}_{\gamma_1}$ and $\mathbf{e}_{\gamma_2}$ have disjoint support:
\begin{equation}
	a_0 = \langle v_0,\, L_\mathrm{sp}\,v_0\rangle = -i\mu\,\langle \mathbf{e}_{\gamma_1},\mathbf{e}_{\gamma_2}\rangle = 0.
\end{equation}
With $b_0 = 0$ by convention:
\begin{equation}
	\begin{aligned} w_1 =& -i\mu\,\mathbf{e}_{\gamma_2}, \quad b_1 = \|-i\mu\,\mathbf{e}_{\gamma_2}\| = |\mu|, \\ v_1 =& -i\,\mathbf{e}_{\gamma_2} = (-i)^1\,\mathbf{e}_{\gamma_2} \qquad (\mu > 0). \end{aligned}
	\label{eq:base_result_app}
\end{equation}
This establishes $v_1 = (-i)^1\,\mathbf{e}_{\gamma_2}$ and $b_1 = |\mu|$, consistent with Eq.~\eqref{eq:vn_claim_app} at $n=1$.

\subsubsection*{\texorpdfstring{Inductive step A: odd to even ($v_{2n-1} \to v_{2n}$)}{Inductive step A: odd to even}}

\noindent\textit{Hypothesis.} $v_{2n-1} = (-i)^{2n-1}\,\mathbf{e}_{\gamma_{2n}}$ and $b_{2n-1} = |\mu|$. Since $(-i)^{2n-1} = (-1)^n\,i$ (as $(-i)^{2n}=(-1)^n$ and $(-i)^{-1}=i$), we have $v_{2n-1} = (-1)^n\,i\,\mathbf{e}_{\gamma_{2n}}$ (purely imaginary).

\noindent\textit{Apply $L_\mathrm{sp}$:}
\begin{align}
	L_\mathrm{sp}\,v_{2n-1} =& (-1)^n\,i \cdot L_\mathrm{sp}\,\mathbf{e}_{\gamma_{2n}} \\
	\stackrel{\eqref{eq:Lsp_B_app}}{=}& (-1)^n\,i\,\bigl(+i\mu\,\mathbf{e}_{\gamma_{2n-1}} - 2it\,\mathbf{e}_{\gamma_{2n+1}}\bigr) \notag\\ =& (-1)^n\bigl(i^2\mu\,\mathbf{e}_{\gamma_{2n-1}} - 2i^2 t\,\mathbf{e}_{\gamma_{2n+1}}\bigr) \\
	=& (-1)^n\bigl(-\mu\,\mathbf{e}_{\gamma_{2n-1}} + 2t\,\mathbf{e}_{\gamma_{2n+1}}\bigr),
	\label{eq:Lsp_odd_result_app}
\end{align}
a real vector.

\noindent\textit{Diagonal coefficient:} $a_{2n-1} = \langle v_{2n-1},\, L_\mathrm{sp}\,v_{2n-1}\rangle = 0$ by orthogonality of $\mathbf{e}_{\gamma_{2n}}$ to both $\mathbf{e}_{\gamma_{2n-1}}$ and $\mathbf{e}_{\gamma_{2n+1}}$.

\noindent\textit{Residual.} The previous vector is $v_{2n-2} = (-i)^{2n-2}\,\mathbf{e}_{\gamma_{2n-1}} = (-1)^{n-1}\,\mathbf{e}_{\gamma_{2n-1}}$. Subtracting $b_{2n-1}\,v_{2n-2} = |\mu|\,(-1)^{n-1}\,\mathbf{e}_{\gamma_{2n-1}} = -(-1)^n\mu\,\mathbf{e}_{\gamma_{2n-1}}$:
\begin{equation}
	\begin{aligned}
		w_{2n} =& (-1)^n\bigl(-\mu\,\mathbf{e}_{\gamma_{2n-1}} + 2t\,\mathbf{e}_{\gamma_{2n+1}}\bigr) - \bigl[-(-1)^n\mu\,\mathbf{e}_{\gamma_{2n-1}}\bigr] \\ 
		=& (-1)^n\,(-\mu+\mu)\,\mathbf{e}_{\gamma_{2n-1}} + (-1)^n\,2t\,\mathbf{e}_{\gamma_{2n+1}}\\
		=& (-1)^n\,2t\,\mathbf{e}_{\gamma_{2n+1}},
	\end{aligned}
	\label{eq:w2n_app}
\end{equation}
where the $\mathbf{e}_{\gamma_{2n-1}}$ terms cancel exactly.
\begin{equation}
	\boxed{b_{2n} = 2|t|, \qquad v_{2n} = (-1)^n\,\mathbf{e}_{\gamma_{2n+1}} = (-i)^{2n}\,\mathbf{e}_{\gamma_{2n+1}}.}
	\label{eq:step_A_result_app}
\end{equation}

\subsubsection*{\texorpdfstring{Inductive step B: even to odd ($v_{2n} \to v_{2n+1}$)}{Inductive step B: even to odd}}

\noindent\textit{Hypothesis.} $v_{2n} = (-1)^n\,\mathbf{e}_{\gamma_{2n+1}}$ (real) and $b_{2n} = 2|t|$.

\noindent\textit{Apply $L_\mathrm{sp}$:}
\begin{align}
	L_\mathrm{sp}\,v_{2n} =& (-1)^n\,L_\mathrm{sp}\,\mathbf{e}_{\gamma_{2n+1}} \\
	\stackrel{\eqref{eq:Lsp_A_app}}{=}& (-1)^n\,\bigl(+2it\,\mathbf{e}_{\gamma_{2n}} - i\mu\,\mathbf{e}_{\gamma_{2n+2}}\bigr),
	\label{eq:Lsp_even_result_app}
\end{align}
a purely imaginary vector.

\noindent\textit{Diagonal coefficient:} $a_{2n} = \langle v_{2n},\, L_\mathrm{sp}\,v_{2n}\rangle = 0$ by orthogonality of $\mathbf{e}_{\gamma_{2n+1}}$ to both $\mathbf{e}_{\gamma_{2n}}$ and $\mathbf{e}_{\gamma_{2n+2}}$.

\noindent\textit{Residual.} The previous vector is $v_{2n-1} = (-1)^n\,i\,\mathbf{e}_{\gamma_{2n}}$. Subtracting $b_{2n}\,v_{2n-1} = 2t\cdot(-1)^n\,i\,\mathbf{e}_{\gamma_{2n}} = (-1)^n\,2it\,\mathbf{e}_{\gamma_{2n}}$ (using $|t|=t$):
\begin{equation}
	\begin{aligned}
		w_{2n+1} =& (-1)^n\,\bigl(2it\,\mathbf{e}_{\gamma_{2n}} - i\mu\,\mathbf{e}_{\gamma_{2n+2}}\bigr) - (-1)^n\,2it\,\mathbf{e}_{\gamma_{2n}} \\
		=& (-1)^n\,(2it-2it)\,\mathbf{e}_{\gamma_{2n}} + (-1)^n\,(-i\mu)\,\mathbf{e}_{\gamma_{2n+2}} \\
		=& (-1)^{n+1}\,i\mu\,\mathbf{e}_{\gamma_{2n+2}},
	\end{aligned}
	\label{eq:w2np1_app}
\end{equation}
where the $\mathbf{e}_{\gamma_{2n}}$ terms cancel exactly, and $(-1)^n\,(-1) = (-1)^{n+1}$.
\begin{equation}
	\boxed{b_{2n+1} = |\mu|, \quad v_{2n+1} = (-1)^{n+1}\,i\,\mathbf{e}_{\gamma_{2n+2}} = (-i)^{2n+1}\,\mathbf{e}_{\gamma_{2n+2}}.}
	\label{eq:step_B_result_app}
\end{equation}
The last equality uses $(-i)^{2n+1} = (-i)^{2n}\cdot(-i) = (-1)^n\cdot(-i) = (-1)^{n+1}\,i$.

Steps A and B propagate the induction for all $n$, establishing $v_n = (-i)^n\,\mathbf{e}_{\gamma_{n+1}}$, $b_{2n-1}=|\mu|$, $b_{2n}=2|t|$, and $a_n=0$ for all $n$. \hfill$\square$

\paragraph{Why the cancellations occur.}
At $\Delta = t$, $\mathcal{H}_M$ is tridiagonal: $L_\mathrm{sp}$ maps each basis vector $\mathbf{e}_{\gamma_\mu}$ to a combination of at most two neighbors $\mathbf{e}_{\gamma_{\mu-1}}$ and $\mathbf{e}_{\gamma_{\mu+1}}$. One neighbor is always proportional to $v_{n-1}$; the Gram--Schmidt subtraction removes it exactly, leaving a single-component residual $w_{n+1}$ that defines $v_{n+1}$. The overall phase factor $(-i)^n$ has unit magnitude and propagates through every step without affecting any norm or inner product. At $\Delta \neq t$, the entry $\mathcal{H}_{M,\,2j-1,\,2j+2} = +(\Delta-t) \neq 0$ introduces next-nearest-neighbor connections in $\mathcal{H}_M$; $L_\mathrm{sp}\,\mathbf{e}_{\gamma_\mu}$ then acquires a component at distance two, the Krylov vectors become superpositions of multiple basis vectors, and the alternating flat structure is broken.

\subsection{SSH interpretation and topological criterion}
\label{app:srk_ssh}

\paragraph{\texorpdfstring{The Su--Schrieffer--Heeger (SSH) model.}{The Su-Schrieffer-Heeger (SSH) model}}
The SSH model~\cite{Su1979, Asboth} is the simplest 1D tight-binding chain with a topological phase. Label the two sublattice sites within unit cell $j$ as $A_j$ (left) and $B_j$ (right), with $c_{j,A}$ and $c_{j,B}$ the corresponding fermionic annihilation operators. The Hamiltonian on an open chain of $L$ unit cells is
\begin{equation}
	H_{\mathrm{SSH}}
	= t_1 \sum_{j=1}^{L}
	\bigl(c_{j,A}^{\dagger}\,c_{j,B}^{\phantom{\dagger}} + \mathrm{h.c.}\bigr)
	+ t_2 \sum_{j=1}^{L-1}
	\bigl(c_{j,B}^{\dagger}\,c_{j+1,A}^{\phantom{\dagger}} + \mathrm{h.c.}\bigr),
	\label{eq:H_ssh_app}
\end{equation}
where $t_1$ is the hopping \emph{within} a unit cell (between $A_j$ and $B_j$ on adjacent physical sites; intracell) and $t_2$ is the hopping \emph{between} neighboring cells (between $B_j$ and $A_{j+1}$; intercell).
With open boundary conditions the chain has two phases:
\begin{itemize}
	\item \emph{Trivial} ($t_1 > t_2$): every $A_j$ pairs strongly with $B_j$ inside its own unit cell; no site is left unbonded at the boundary.
	\item \emph{Topological} ($t_2 > t_1$): the strong dimers are instead $B_j$--$A_{j+1}$ \emph{across} cell boundaries; the leftmost site $A_1$ and the rightmost site $B_L$ have no strong partner and become exponentially localized zero-energy edge modes.
\end{itemize}
The critical point $t_1 = t_2$ separates the two phases, and the gap closes there. No continuous path between the two phases exists without closing the gap (for the symmetry class of the SSH model), which is the defining signature of a topological phase transition.

\paragraph{\texorpdfstring{The Majorana matrix $\mathcal{H}_M$ is an SSH Hamiltonian.}{The Majorana matrix is an SSH Hamiltonian.}}
Recall from Eq.~\eqref{eq:HM_tridiag_app} that at $\Delta = t$ the only nonzero entries of $\mathcal{H}_M$ are
\begin{equation}
	\begin{aligned}
		\mathcal{H}_{M,\,2j-1,\,2j} =& +\mu \quad\text{(intra-site)}, \\
		\mathcal{H}_{M,\,2j,\,2j+1} =& +2t \quad\text{(inter-site)},
	\end{aligned}
	\label{eq:HM_ssh_entries_app}
\end{equation}
making $\mathcal{H}_M$ a real antisymmetric tridiagonal matrix with strictly alternating entries $|\mu|$ and $2|t|$. Because the single-particle Liouvillian $L_{\mathrm{sp}} = i\mathcal{H}_M$ is Hermitian, its eigenvalue equation is equivalent to a standard tight-binding problem. Concretely, make the identification
\begin{equation}
	A_j \;\leftrightarrow\; \gamma_{2j-1}, \quad
	B_j \;\leftrightarrow\; \gamma_{2j}, \quad
	t_1 \;\leftrightarrow\; |\mu|, \quad
	t_2 \;\leftrightarrow\; 2|t|.
	\label{eq:ssh_map_app}
\end{equation}
Under this identification \(L_{\rm sp}\) and \(H_{\rm SSH}\) are the \textit{same matrix}: the spectrum of \(L_{\rm sp}\) is identical to the single-particle spectrum of \(H_{\rm SSH}(t_1=\mu,t_2=2t)\). Equivalently, at \(\Delta=t\) the balanced short-range Kitaev chain, written in Majorana variables, is exactly the SSH chain at the single-particle level, and the Krylov construction does not deform this matrix but simply reproduces its alternating-bond structure. This is not an analogy but an exact algebraic equality, confirmed numerically to machine precision.

The pictorial chain
\begin{equation}
	(\gamma_1 \xleftrightarrow{|\mu|} \gamma_2)
	\xleftrightarrow{2|t|}
	(\gamma_3 \xleftrightarrow{|\mu|} \gamma_4)
	\xleftrightarrow{2|t|}
	(\gamma_5 \xleftrightarrow{|\mu|} \gamma_6)
	\xleftrightarrow{2|t|}
	\cdots
	\label{eq:ssh_chain_picture_app}
\end{equation}
is precisely the SSH chain~\eqref{eq:H_ssh_app} with the pairing~\eqref{eq:ssh_map_app}.

\paragraph{Topological criterion from SSH physics.}
The SSH chain~\eqref{eq:H_ssh_app} is topological whenever the \emph{intercell} bond dominates, $t_2 > t_1$. Substituting the identification~\eqref{eq:ssh_map_app}:
\begin{equation}
	t_2 > t_1
	\;\Longleftrightarrow\;
	2|t| > |\mu|
	\;\Longleftrightarrow\;
	|\mu| < 2|t|,
	\label{eq:topo_crit_ssh_app}
\end{equation}
which is exactly the standard Kitaev chain phase boundary~\cite{Kitaev2001}.
When $|\mu| < 2|t|$, the inter-site bonds $B_j\text{--}A_{j+1}$ dominate and $\gamma_1$ (the $A_1$ site) is left without a strong partner, producing a Majorana zero mode localized at the left edge. When $|\mu| > 2|t|$, the intra-site bonds $A_j\text{--}B_j$ dominate and both boundary Majoranas are absorbed into the bulk dimers, leaving no edge mode: the chain is trivial.
The critical point $|\mu| = 2|t|$ is the gap-closing transition.

\paragraph{The Krylov chain inherits the SSH structure.}
The induction proof of Sec.~\ref{app:srk_proof} establishes that the Krylov basis vectors satisfy $v_n = (-i)^n \mathbf{e}_{\gamma_{n+1}}$, so the recursion \emph{traverses the Majorana chain one site at a time}, reading off bond strengths sequentially. The resulting Krylov tridiagonal (the operator Krylov matrix, whose off-diagonals are the Lanczos coefficients) is therefore
\begin{equation}
	v_0 \xleftrightarrow{b_1} v_1 \xleftrightarrow{b_2} v_2
	\xleftrightarrow{b_3} v_3 \xleftrightarrow{b_4} v_4 \cdots,
	(b_{2n-1} = |\mu|, b_{2n} = 2|t|),
	\label{eq:krylov_ssh_app}
\end{equation}
which is structurally identical to the SSH chain~\eqref{eq:H_ssh_app} with the same bond strengths. The three chains---real-space Majorana chain, SSH tight-binding chain, and Krylov chain---are therefore related by the exact dictionary:
\begin{center}
	\renewcommand{\arraystretch}{1.35}
	\begin{tabular}{lccc}
		\hline
		& Majorana chain & SSH chain & Krylov chain \\
		\hline
		Left sublattice  & $\gamma_{2j-1}$ & $A_j$  & $v_{2j-2}$ \\
		Right sublattice & $\gamma_{2j}$   & $B_j$  & $v_{2j-1}$ \\
		Intracell bond   & $|\mu|$         & $t_1$  & $b_{2n-1}$ \\
		Intercell bond   & $2|t|$          & $t_2$  & $b_{2n}$   \\
		\hline
	\end{tabular}
\end{center}

\paragraph{Staggering parameter as the SSH order parameter.}
The Krylov staggering parameter $\eta_n = \ln(b_{2n-1}/b_{2n})$ measures the \emph{logarithmic ratio of weak to strong bond}, exactly as the SSH dimerization ratio $\ln(t_1/t_2)$ does. Its sign is therefore a sharp topological indicator:
\begin{equation}
	\eta = \ln\!\left(\frac{|\mu|}{2|t|}\right)
	\begin{cases}
		< 0 & |\mu| < 2|t| \\
		\text{ } &  \text{(topological: intercell bond dominates)}, \\
		= 0 & |\mu| = 2|t|  \\
		\text{ }  &  \text{(critical point: gap closes)}, \\
		> 0 & |\mu| > 2|t|  \\
		\text{ }  &  \text{(trivial: intracell bond dominates)}.
	\end{cases}
	\label{eq:eta_ssh_app}
\end{equation}
Because $\eta_n$ is $n$-independent and analytic in $\mu$ and $t$, it provides a size-independent (all the way up to thermodynamic limit), finite-$n$ diagnostic of the topological phase that requires no finite-size extrapolation.

\paragraph{Important caveat.}
The equivalence~\eqref{eq:ssh_map_app} holds at the level of the \emph{single-particle Majorana matrix} $L_{\mathrm{sp}} = i\mathcal{H}_M$, not of the full many-body Hamiltonian. The physical system is a $p$-wave superconductor (Kitaev chain), not an insulator (SSH chain); the two differ in their symmetry classes, ground-state particle-number conservation, and many-body structure. The precise statement is: at $\Delta = t$ and with the boundary-Majorana seed $\gamma_1$, the Majorana matrix and the Krylov matrix are both alternating-bond 1D chains with the same bond strengths, so the Krylov topological diagnostic $\eta < 0$ and the Kitaev phase criterion $|\mu| < 2|t|$ agree exactly. This agreement is a structural consequence of the balanced hopping-pairing condition $\Delta = t$, which eliminates all next-nearest-neighbor bonds from $\mathcal{H}_M$ and makes the chain maximally local.

\subsection{Physical interpretation and topological criterion}
\label{app:srk_physics}

\begin{figure}[tb]
	\centering
    \begin{overpic}{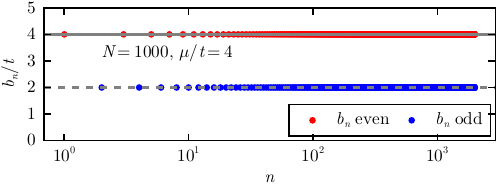}
        \put(14.5,26){(a)}
    \end{overpic}
    \begin{overpic}{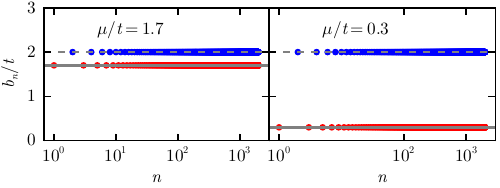}
        \put(13,30.5){(b)}
        \put(58.5,30.5){(c)}
    \end{overpic}
    \begin{overpic}{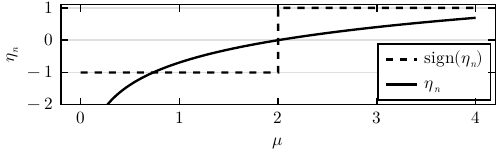}
        \put(14.5,24){(d)}
    \end{overpic}
    {\phantomsubcaption\label{fig:app_exact_srk_bn_mu=4}}
    {\phantomsubcaption\label{fig:app_exact_srk_bn_mu=1.7}}
    {\phantomsubcaption\label{fig:app_exact_srk_bn_mu=0.3}}
    {\phantomsubcaption\label{fig:app_exact_srk_eta_mu}}
    \caption{
        Numerical verification of Eqs.~\eqref{eq:exact_bn_app}--\eqref{eq:eta_exact_app} for the balanced short-range Kitaev chain at $|t| = |\Delta| = 1$ ($\epsilon = 0$), with seed $\gamma_1$ and $N = 1000$ sites. \subref{fig:app_exact_srk_bn_mu=4}~Off-diagonal Lanczos coefficients $b_n$ at $\mu/t = 4$ (trivial regime). \subref{fig:app_exact_srk_bn_mu=1.7},\subref{fig:app_exact_srk_bn_mu=0.3}~Off-diagonal Lanczos coefficients $b_n$ at $\mu/t = 1.7$ and $\mu/t = 0.3$ (topological regime).
        In all three cases, the numerical coefficients match the exact alternating values set by $|\mu|$ and $2|t|$ which are denoted by a solid gray line and dashed gray line, respectively.
        Lanczos coefficients are obtained using the same numerical procedure as in \cref{fig:lanczos_coeffs_maj1}.
        \subref{fig:app_exact_srk_eta_mu}~Staggering parameter $\eta = \ln(b_{2n-1}/b_{2n}) = \ln(|\mu|/2|t|)$ as a function of $\mu$ at fixed $t = 1$; its sign changes at $|\mu| = 2|t|$, reproducing the topological phase boundary. Numerical and analytical values agree to machine precision.
    }
	\label{fig:app_exact_srk}
\end{figure}

The Krylov basis vectors traverse the Majorana chain~\eqref{eq:chain_picture_app} sequentially, and $b_n$ records the bond strength encountered at each step: $|\mu|$ for the intra-site (A--B) bond and $2|t|$ for the inter-site (B--A) bond. Because these two values alternate with strict period two for every $\mu$, $t$, and $N$, the staggering parameter defined in Eq.~\eqref{eq:etadef_main} is the $n$-independent constant
\begin{equation}
	\eta_n = \ln\!\left(\frac{|\mu|}{2|t|}\right) = \mathrm{const.}, \qquad \forall\, n,
	\label{eq:eta_const_app}
\end{equation}
whose sign reproduces the SSH topological criterion~\eqref{eq:topo_crit_ssh_app} and, equivalently, the standard Kitaev chain phase boundary $|\mu| = 2|t|$ at $t = \Delta$~\cite{Kitaev2001,Alicea2012}. Because Eq.~\eqref{eq:eta_const_app} is exact for every finite \(N\) and every value of \(n\), it requires no thermodynamic extrapolation and no finite-size correction: the staggering parameter \(\eta_n\) is an exact, size-independent, analytic topological indicator for the short-range Kitaev chain at \(\epsilon = 0\). Numerical confirmation at \(N = 1000\) is shown in Fig.~\ref{fig:app_exact_srk}, where panels (a)--(c) verify the exact alternating Lanczos coefficients for representative trivial and topological points \(\mu/t = 4,\,1.7,\,0.3\), and panel (d) shows that \(\eta = \ln(|\mu|/2|t|)\) changes sign at \(|\mu| = 2|t|\), reproducing the topological phase boundary.

	\begin{figure*}[t]
		\centering
		\begin{overpic}{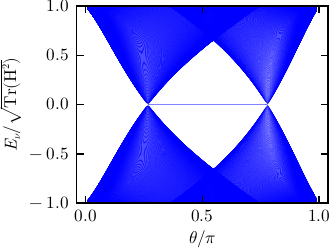}
			\put(32,66){\color{white}(a) $\alpha=3$}
		\end{overpic}
		\begin{overpic}{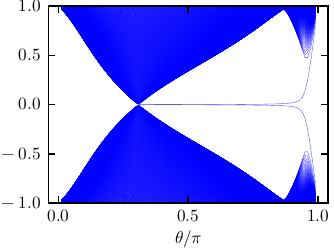}
			\put(27,66){\color{white}(b) $\alpha=1$}
		\end{overpic}
		\begin{overpic}{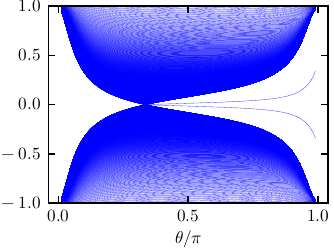}
			\put(27,66){\color{white}(c) $\alpha=1/3$}
		\end{overpic}
		
		{\phantomsubcaption\label{fig:spectrum_alpha30_eps_1}}
		{\phantomsubcaption\label{fig:spectrum_alpha10_eps_1}}
		{\phantomsubcaption\label{fig:spectrum_alpha013_eps_1}}
		\caption{
			BdG spectrum $E_\nu/\sqrt{\mathrm{Tr}(H^2)}$ versus $\theta/\pi$ at $\epsilon=1$ for three long-range exponents $\alpha$ ($N=1000$, open boundaries).
			The $\alpha$-dependence mirrors \cref{fig:spectrum_eps_minus02}: panel~(a) displays a gapless region at $\theta/\pi \approx 0.5$ with sparse spectral density near $E=0$ elsewhere; panel~(b) shows modified spectral density; panel~(c) shows the degeneracy lifted throughout the $\theta/\pi$ interval except for a small near-degenerate regime, with dense spectral density near $E=0$.
			Compared to $\epsilon=-0.2$, the spectral density near $E=0$ is increased across all $\alpha$ regimes.
		}
		\label{fig:spectrum_eps_1}
	\end{figure*}
	
	\begin{figure*}[htbp]
		\centering			
		\begin{overpic}{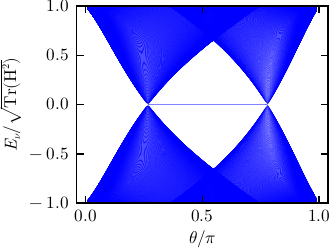}
			\put(32,66){\color{white}(a) $\alpha=3$}
		\end{overpic}
		\begin{overpic}{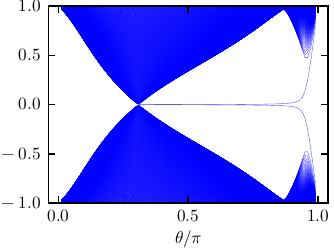}
			\put(27,66){\color{white}(b) $\alpha=1$}
		\end{overpic}
		\begin{overpic}{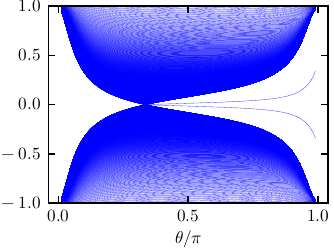}
			\put(27,66){\color{white}(c) $\alpha=1/3$}
		\end{overpic}
		{\phantomsubcaption\label{fig:spectrum_alpha30_eps_10}}
		{\phantomsubcaption\label{fig:spectrum_alpha10_eps_10}}
		{\phantomsubcaption\label{fig:spectrum_alpha013_eps_10}}
		
		\caption{%
			BdG spectrum $E_\nu/\sqrt{\mathrm{Tr}(H^2)}$ versus $\theta/\pi$ at extreme pairing-dominated $\epsilon=10$ for three long-range exponents $\alpha$ ($N=1000$, open boundaries).
			The systematic $\alpha$-dependence observed in Figs.~\ref{fig:spectrum_eps_minus02} and~\ref{fig:spectrum_eps_1} persists: gapless region for $\alpha=3$, similar gapless range for $\alpha=1$ with gap opening, and degeneracy lifted except for a small near-degenerate regime for $\alpha=1/3$.
			Spectral density near $E=0$ is further increased relative to smaller $\epsilon$, occurring uniformly across all $\alpha$ including the short-range limit.
			The qualitative $\alpha$-dependence remains unchanged across various $\epsilon$ regimes.
		}
		\label{fig:spectrum_eps_10}
	\end{figure*}

	\section{Additional Results for Spectrum and the Role of Long-Range Pairing}
	\label{app:spectra_additional_epsilon}
	
	To establish the robustness of the spectral trends documented in Sec.~\ref{subsec:spectrum_impact_lr}, we present here the BdG spectra for $\epsilon=1$ and $\epsilon=10$, which span from moderate to extreme pairing dominance relative to hopping.

	Figure~\ref{fig:spectrum_eps_1} displays the spectrum at $\epsilon=1$.
	The qualitative $\alpha$-dependence observed at $\epsilon=-0.2$ (Fig.~\ref{fig:spectrum_eps_minus02}) persists: panel~\subref{fig:spectrum_alpha30_eps_1} ($\alpha=3$) exhibits a gapless region centered near $\theta/\pi \approx 0.5$ with sparse spectral density near $E=0$ elsewhere; panel~\subref{fig:spectrum_alpha10_eps_1} ($\alpha=1$) shows modified spectral density; and panel~\subref{fig:spectrum_alpha013_eps_1} ($\alpha=0.1$) displays the degeneracy lifted except for a small near-degenerate regime, with markedly denser spectral density near $E=0$ throughout.
	Compared to $\epsilon=-0.2$, the enhanced pairing amplitude at $\epsilon=1$ increases the density of modes near $E=0$ across all three $\alpha$ values.
	
	Figure~\ref{fig:spectrum_eps_10} presents the extreme pairing-dominated limit $\epsilon=10$.
	The systematic $\alpha$-dependence remains qualitatively unchanged: gapless region over a finite $\theta/\pi$ range for $\alpha=3$ (panel~\subref{fig:spectrum_alpha30_eps_10}), similar gapless range with gap opening for $\alpha=1$ (panel~\subref{fig:spectrum_alpha10_eps_10}), and degeneracy lifting to a small near-degenerate regime for $\alpha=1/3$ (panel~\subref{fig:spectrum_alpha013_eps_10}).
	The spectral density near $E=0$ is further amplified relative to both $\epsilon=-0.2$ and $\epsilon=1$, reflecting the dominance of the pairing term $(1+\epsilon)\sin\theta$ in Eq.~\eqref{eq:Delta_def}.
	Notably, this increased density occurs uniformly across all $\alpha$ regimes, including the short-range limit where exponential localization would otherwise suppress accumulation of states near $E=0$.

	These results confirm that the qualitative $\theta$-dependence and the trends in the density of low-energy BdG modes near $E_0$, which underlie the analysis in Sec.~\ref{sec:results}, persist across different hopping–pairing ratios.
	The systematic increase in the density of low-energy BdG modes near $E_0$ with $\epsilon$ provides an additional control parameter, but does not alter the fundamental $\alpha$-dependence of the eigenmode structure.

	\begin{figure*}[htbp]
		\centering
		\begin{overpic}{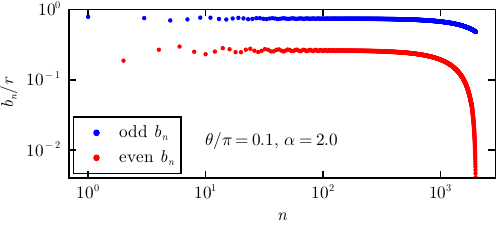}
			\put(43,22){(a) bulk gap}
		\end{overpic}
		\begin{overpic}{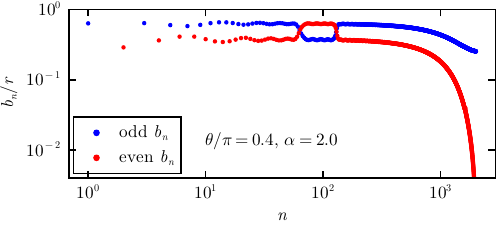}
			\put(43,22){(c) edge gap}
		\end{overpic}
		\vspace{2mm}
		\begin{overpic}{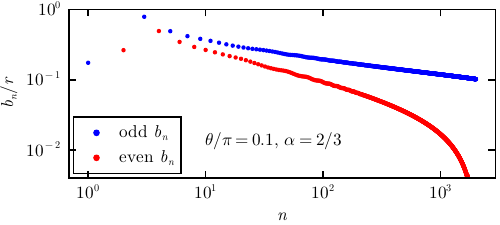}
			\put(43,22){(b) bulk gap}
		\end{overpic}
		\begin{overpic}{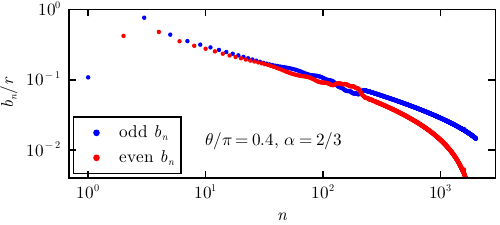}
			\put(43,22){(d) edge gap}
		\end{overpic}
		
		{\phantomsubcaption\label{fig:lanczos_coefficients_maj12_shortrange_trivial}}
		{\phantomsubcaption\label{fig:lanczos_coefficients_maj12_shortrange_edge}}
		{\phantomsubcaption\label{fig:lanczos_coefficients_maj12_longrange_trivial}}
		{\phantomsubcaption\label{fig:lanczos_coefficients_maj12_longrange_edge}}
		\caption{
			Lanczos coefficients $\{b_n\}$ for the long-range Kitaev chain at $\epsilon=-0.2$ with open boundaries and boundary seed $\gamma_1 + \gamma_2$ ($N=1000$).
			Each panel shows the two interleaved subsequences (odd and even recursion steps), whose relative ordering determines the staggering parameter $\eta_n=\ln(b_{2n-1}/b_{2n})$ and crossing count $N_{\mathrm{cross}}$.
			Panels \subref{fig:lanczos_coefficients_maj12_shortrange_trivial} ($\alpha=2$, $\theta/\pi=0.1$) and \subref{fig:lanczos_coefficients_maj12_longrange_trivial} ($\alpha=2/3$, $\theta/\pi=0.1$) lie in the bulk-gap regime and exhibit no interchange of the two subsequences ($N_{\mathrm{cross}}=0$), while panels \subref{fig:lanczos_coefficients_maj12_shortrange_edge} ($\alpha=2$, $\theta/\pi=0.4$) and \subref{fig:lanczos_coefficients_maj12_longrange_edge} ($\alpha=2/3$, $\theta/\pi=0.4$) lie in the edge-gap regime and show clear interchanges ($N_{\mathrm{cross}}\ge 1$).
			The qualitative pattern of no crossing in the bulk-gap phase as well as nonzero crossings in the edge-gap phase matches that observed for the single-operator seed $\gamma_1$ (Fig.~\ref{fig:lanczos_coeffs_maj1}), though the quantitative sharpness of the phase diagram when scanned over full parameter space $(\alpha, \theta)$ is reduced compared to the minimal boundary seed presented in the main text (compare Fig.~\ref{fig:lrk_edge_bulk_phase_diagram} and Fig.~\ref{fig:lrk_edge_bulk_phase_diagram_majs}). As in Fig.~\ref{fig:lanczos_coeffs_maj1}, the total number of Lanczos coefficients varies across parameter points. To ensure numerical stability, the recursion is terminated when $b_n \lesssim 10^{-7}$, and all subsequent coefficients are excluded from the analysis (see Appendix~\ref{app:alg_numerical_stability} for details on stability checks).
		}
		\label{fig:lanczos_coeffs_maj12}
	\end{figure*}

	\section{Alternative seed operators}
	\label{app:alt_seeds}
	
	To assess robustness with respect to the choice of local probe, we also consider the following Hermitian linear operators in the Majorana basis: 
	\begin{align}
		\mathcal{O}_{\mathrm{edge,2}}(0) &= \gamma_1+\gamma_2\\
		\mathcal{O}_{\mathrm{mid}}(0) &= \gamma_{N},\\
		\mathcal{O}_{\mathrm{mid},2}(0) &= \gamma_{N}+\gamma_{N+1}\,.
	\end{align}
	Here $\gamma_\mu$ with $\mu=1,\ldots,2N$ denote the $2N$ Majorana modes of an $N$-site chain.
	All three seeds are Hermitian, therefore their Lanczos tridiagonalization also satisfies $a_n=0$ for all $n$ by Appendix~\ref{app:alg_an_zero}.
	
\begin{figure*}[htbp]
		\centering
		\begin{overpic}{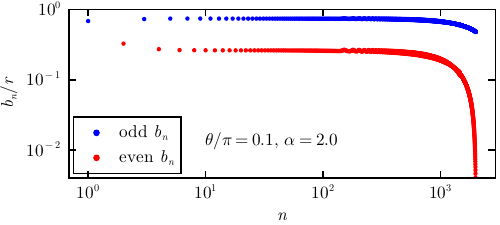}
			\put(43,22){(a) bulk gap}
		\end{overpic}
		\begin{overpic}{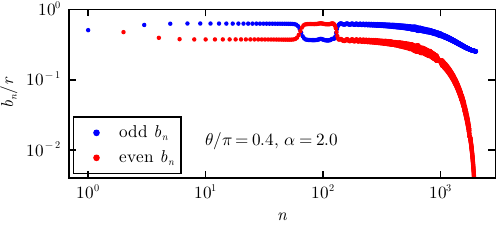}
			\put(43,22){(c) edge gap}
		\end{overpic}
		\vspace{2mm}
		\begin{overpic}{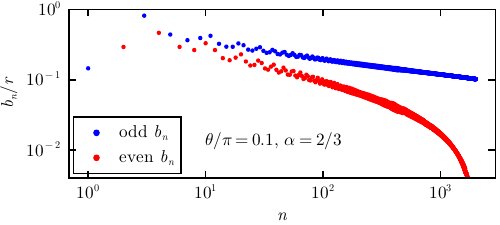}
			\put(43,22){(b) bulk gap}
		\end{overpic}
		\begin{overpic}{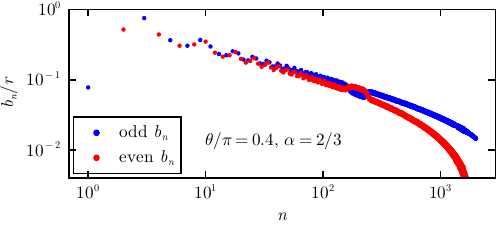}
			\put(43,22){(d) edge gap}
		\end{overpic}
		
		{\phantomsubcaption\label{fig:lanczos_coefficients_majhalf_shortrange_trivial}}
		{\phantomsubcaption\label{fig:lanczos_coefficients_majhalf_shortrange_edge}}
		{\phantomsubcaption\label{fig:lanczos_coefficients_majhalf_longrange_trivial}}
		{\phantomsubcaption\label{fig:lanczos_coefficients_majhalf_longrange_edge}}
		\caption{%
			Lanczos coefficients $\{b_n\}$ for the long-range Kitaev chain at $\epsilon=-0.2$ with open boundaries and bulk seed $\gamma_N$ ($N=1000$).
			Each panel shows the two interleaved subsequences (odd and even recursion steps), whose relative ordering determines the staggering parameter $\eta_n=\ln(b_{2n-1}/b_{2n})$ and crossing count $N_{\mathrm{cross}}$.
			Panels \subref{fig:lanczos_coefficients_majhalf_shortrange_trivial} ($\alpha=2$, $\theta/\pi=0.1$) and \subref{fig:lanczos_coefficients_majhalf_longrange_trivial} ($\alpha=2/3$, $\theta/\pi=0.1$) lie in the bulk-gap regime ($N_{\mathrm{cross}}=0$), while panels \subref{fig:lanczos_coefficients_majhalf_shortrange_edge} ($\alpha=2$, $\theta/\pi=0.4$) and \subref{fig:lanczos_coefficients_majhalf_longrange_edge} ($\alpha=2/3$, $\theta/\pi=0.4$) lie in the edge-gap regime ($N_{\mathrm{cross}}\ge 1$).
			The qualitative edge-bulk distinction persists for this bulk-localized seed, which couples comparably to both chain ends, though the phase diagram (obtained by scanning the full parameter space $(\alpha, \theta)$) sharpness degrades compared to boundary seeds (compare Fig.~\ref{fig:lrk_edge_bulk_phase_diagram} and Fig.~\ref{fig:lrk_edge_bulk_phase_diagram_majs}). To maintain numerical rigor, the recursion is terminated when $b_n \lesssim 10^{-7}$, excluding all subsequent coefficients (see Appendix~\ref{app:alg_numerical_stability}).
		}
		\label{fig:lanczos_coeffs_majhalf}
	\end{figure*}

\begin{figure*}[htbp]
		\centering
		\begin{overpic}{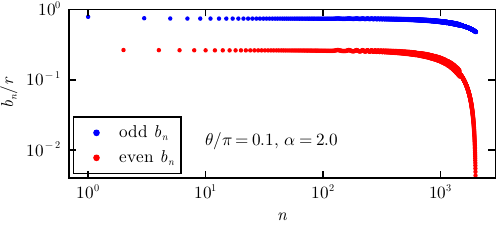}
			\put(43,22){(a) bulk gap}
		\end{overpic}
		\begin{overpic}{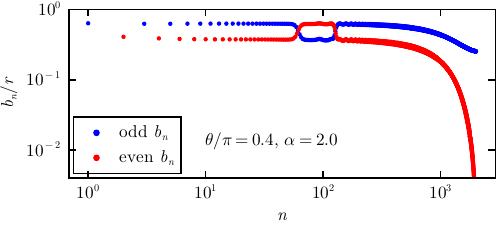}
			\put(43,22){(c) edge gap}
		\end{overpic}
		\vspace{2mm}
		\begin{overpic}{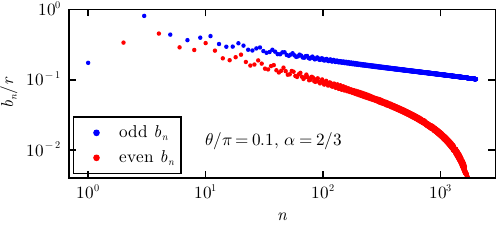}
			\put(43,22){(b) bulk gap}
		\end{overpic}
		\begin{overpic}{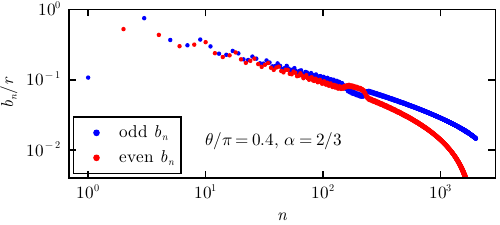}
			\put(43,22){(d) edge gap}
		\end{overpic}
		
		{\phantomsubcaption\label{fig:lanczos_coefficients_majhalf12_shortrange_trivial}}
		{\phantomsubcaption\label{fig:lanczos_coefficients_majhalf12_shortrange_edge}}
		{\phantomsubcaption\label{fig:lanczos_coefficients_majhalf12_longrange_trivial}}
		{\phantomsubcaption\label{fig:lanczos_coefficients_majhalf12_longrange_edge}}
		\caption{%
			Lanczos coefficients $\{b_n\}$ for the long-range Kitaev chain at $\epsilon=-0.2$ with open boundaries and bulk seed $\gamma_N + \gamma_{N+1}$ ($N=1000$).
			Each panel shows the two interleaved subsequences whose relative ordering determines the staggering parameter $\eta_n=\ln(b_{2n-1}/b_{2n})$ and crossing count $N_{\mathrm{cross}}$.
			Panels \subref{fig:lanczos_coefficients_majhalf12_shortrange_trivial}, \subref{fig:lanczos_coefficients_majhalf12_longrange_trivial} ($\alpha=2, 2/3$; $\theta/\pi=0.1$) show bulk-gap regime ($N_{\mathrm{cross}}=0$), while panels \subref{fig:lanczos_coefficients_majhalf12_shortrange_edge}, \subref{fig:lanczos_coefficients_majhalf12_longrange_edge} ($\alpha=2, 2/3$; $\theta/\pi=0.4$) show edge-gap regime ($N_{\mathrm{cross}}\ge 1$).
			As with $\gamma_N$, the edge-bulk pattern remains qualitatively visible, though the phase diagram obtained by scanning the full parameter space $(\alpha, \theta)$ shows degraded 
			sharpness compared to boundary seeds (compare Fig.~\ref{fig:lrk_edge_bulk_phase_diagram} and Fig.~\ref{fig:lrk_edge_bulk_phase_diagram_majs}). The recursion depth varies due to our stability criterion $b_n \lesssim 10^{-7}$, beyond which all coefficients are excluded (Appendix~\ref{app:alg_numerical_stability}).
		}
		\label{fig:lanczos_coeffs_majhalf12}
	\end{figure*}

	Figures~\ref{fig:lanczos_coeffs_maj12}, \ref{fig:lanczos_coeffs_majhalf}, and \ref{fig:lanczos_coeffs_majhalf12} show the Lanczos coefficients for these three seeds at representative parameter points spanning both short-range and long-range regimes, as well as bulk-gap and edge-gap phases. 
	In all cases, the qualitative pattern observed in the main text persists: the two interleaved subsequences (odd and even recursion steps) do not interchange in the bulk-gap regime, yielding $N_{\mathrm{cross}}=0$, while clear interchanges occur in the edge-gap regime, producing $N_{\mathrm{cross}}\ge 1$. 
	This confirms that the crossing-count signature is robust across seed choices and reflects genuine edge-bulk physics rather than an artifact of the specific seed operator.
	
\begin{figure*}[htbp]
		\centering
		\begin{overpic}{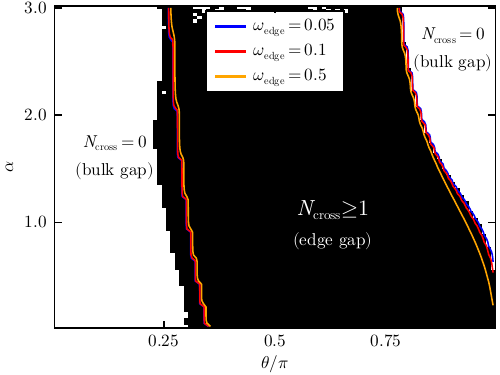}
			\put(65,15){\color{white}(a) $\gamma_1 + \gamma_2$}
		\end{overpic}
		\begin{overpic}{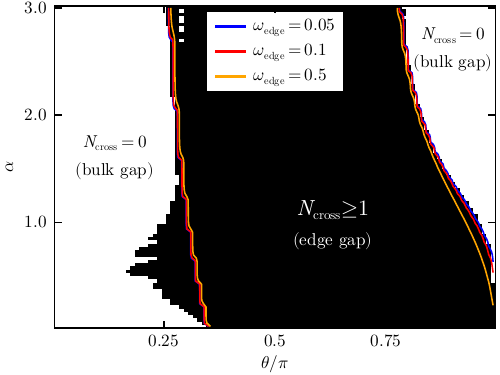}
			\put(70,15){\color{white}(b) $\gamma_N$}
		\end{overpic}
		\begin{overpic}{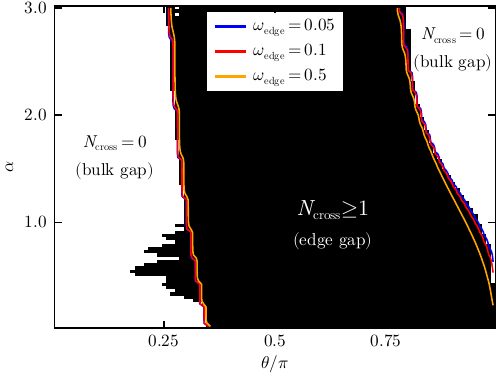}
			\put(65,15){\color{white}(c) $\gamma_N + \gamma_{N+1}$}
		\end{overpic}
		{\phantomsubcaption\label{fig:lrk_edge_bulk_phase_diagram_maj12}}
		{\phantomsubcaption\label{fig:lrk_edge_bulk_phase_diagram_majhalf}}
		{\phantomsubcaption\label{fig:lrk_edge_bulk_phase_diagram_majhalf12}}
		\caption{
			Phase diagrams for the long-range Kitaev chain at $\epsilon=-0.2$ with open boundaries ($N=1000$) showing the crossing-count diagnostic $N_{\mathrm{cross}}(\alpha,\theta)$ (black: $N_{\mathrm{cross}}\ge 1$; white: $N_{\mathrm{cross}}=0$) for different seed operators: \subref{fig:lrk_edge_bulk_phase_diagram_maj12} $\gamma_1 + \gamma_2$ (boundary seed), \subref{fig:lrk_edge_bulk_phase_diagram_majhalf} $\gamma_N$ (bulk seed), \subref{fig:lrk_edge_bulk_phase_diagram_majhalf12} $\gamma_N + \gamma_{N+1}$ (bulk seed). 
			Solid curves show the edge-bulk gap boundary $\Delta_{\mathrm{edge}}=\Delta_{\mathrm{bulk}}$ from the BdG spectrum for three edge-weight thresholds: $\omega_{\mathrm{edge}}=0.05, 0.1, 0.5$ (with $\ell_{\mathrm{edge}}=\lfloor\sqrt{N}\rfloor$).
			The boundary seed \subref{fig:lrk_edge_bulk_phase_diagram_maj12} shows the most quantitative agreement with the BdG-derived boundary (up to grid precision and finite-size), while bulk seeds \subref{fig:lrk_edge_bulk_phase_diagram_majhalf}, \subref{fig:lrk_edge_bulk_phase_diagram_majhalf12} couple to both edges and probe predominantly bulk-extended excitations, resulting in degraded quantitative matching even though qualitative edge-bulk trends remain visible.
			This demonstrates that boundary seeds are essential for obtaining a robust and quantitative Krylov diagnostic of edge-versus-bulk gap control.
			Data computed on a $99\times 99$ grid in $(\alpha,\theta)$ with $\alpha\in(0,3]$ and $\theta\in(0,\pi)$.
		}
		\label{fig:lrk_edge_bulk_phase_diagram_majs}
	\end{figure*}

	However, when the crossing-count diagnostic is computed over the full parameter space $(\alpha,\theta)$, the quantitative sharpness of the phase boundary varies systematically with seed localization, as shown in Fig.~\ref{fig:lrk_edge_bulk_phase_diagram_majs}. 
	The boundary seed $\gamma_1+\gamma_2$ (panel~\subref{fig:lrk_edge_bulk_phase_diagram_maj12}) produces a phase diagram in close quantitative agreement with the BdG-derived edge-bulk gap boundary, though slightly less sharp than the minimal seed $\gamma_1$ used in the main text. 
	By contrast, the bulk seeds $\gamma_N$ and $\gamma_N+\gamma_{N+1}$ (panels~\subref{fig:lrk_edge_bulk_phase_diagram_majhalf}, \subref{fig:lrk_edge_bulk_phase_diagram_majhalf12}) couple comparably to both chain ends and predominantly probe bulk-extended excitations, resulting in noticeably degraded quantitative matching with the BdG boundary even though the qualitative edge-bulk distinction remains visible. 
	These results demonstrate that boundary-localized seeds are essential for obtaining a quantitatively reliable Krylov diagnostic of edge-versus-bulk gap control, while bulk seeds provide at best a qualitative probe.

\end{document}